\documentclass[journal, 10pt, onecolumn]{IEEEtran}
\usepackage{array}
\usepackage{epsfig,epic,eepic,units}
\usepackage{hyperref}
\usepackage{url}
\usepackage{longtable}
\usepackage{mathrsfs}
\usepackage{multirow}
\usepackage{bigstrut}
\usepackage{amssymb}
\usepackage{graphicx}
\usepackage{amsmath}
\usepackage{cases}
\usepackage{algorithm,algorithmic}
\usepackage{etoolbox}
\usepackage{color}
\usepackage{blkarray}
\usepackage[normalem]{ulem}

\newcommand\nc\newcommand
\nc\bfa{{\boldsymbol a}}\nc\bfA{{\boldsymbol A}}\nc\cA{{\mathcal A}}
\nc\bfb{{\boldsymbol b}}\nc\bfB{{\boldsymbol B}}\nc\cB{{\mathcal B}}
\nc\bfc{{\boldsymbol c}}\nc\bfC{{\boldsymbol C}}\nc\cC{{\mathcal C}}
\nc\sC{{\mathscr C}}
\nc\bfd{{\boldsymbol d}}\nc\bfD{{\boldsymbol D}}\nc\cD{{\mathcal D}}
\nc\bfe{{\boldsymbol e}}\nc\bfE{{\boldsymbol E}}\nc\cE{{\mathcal E}}
\nc\bff{{\boldsymbol f}}\nc\bfF{{\boldsymbol F}}\nc\cF{{\mathcal F}}
\nc\bfg{{\boldsymbol g}}\nc\bfG{{\boldsymbol G}}\nc\cG{{\mathcal G}}
\nc\bfh{{\boldsymbol h}}\nc\bfH{{\boldsymbol H}}\nc\cH{{\mathcal H}}
\nc\bfi{{\boldsymbol i}}\nc\bfI{{\boldsymbol I}}\nc\cI{{\mathcal I}}
\nc\bfj{{\boldsymbol j}}\nc\bfJ{{\boldsymbol J}}\nc\cJ{{\mathcal J}}
\nc\bfk{{\boldsymbol k}}\nc\bfK{{\boldsymbol K}}\nc\cK{{\mathcal K}}
\nc\bfl{{\boldsymbol l}}\nc\bfL{{\boldsymbol L}}\nc\cL{{\mathcal L}}
\nc\bfm{{\boldsymbol m}}\nc\bfM{{\boldsymbol M}}\nc\sM{{\mathscr M}}
\nc\bfn{{\boldsymbol n}}\nc\bfN{{\boldsymbol N}}\nc\cN{{\mathcal N}}
\nc\bfo{{\boldsymbol o}}\nc\bfO{{\boldsymbol O}}\nc\cO{{\mathcal O}}
\nc\bfp{{\boldsymbol p}}\nc\bfP{{\boldsymbol P}}\nc\cP{{\mathcal P}}
\nc\bfq{{\boldsymbol q}}\nc\bfQ{{\boldsymbol Q}}\nc\cQ{{\mathcal Q}}
\nc\bfr{{\boldsymbol r}}\nc\bfR{{\boldsymbol R}}\nc\cR{{\mathcal R}}
\nc\bfs{{\boldsymbol s}}\nc\bfS{{\boldsymbol S}}\nc\cS{{\mathcal S}}
\nc\bft{{\boldsymbol t}}\nc\bfT{{\boldsymbol T}}\nc\cT{{\mathcal T}}
\nc\bfu{{\boldsymbol u}}\nc\bfU{{\boldsymbol U}}\nc\cU{{\mathcal U}}
\nc\bfv{{\boldsymbol v}}\nc\bfV{{\boldsymbol V}}\nc\cV{{\mathcal V}}
\nc\bfw{{\boldsymbol w}}\nc\bfW{{\boldsymbol W}}\nc\cW{{\mathcal W}}
\nc\bfx{{\boldsymbol x}}\nc\bfX{{\boldsymbol X}}\nc\cX{{\mathcal X}}
\nc\bfy{{\boldsymbol y}}\nc\bfY{{\boldsymbol Y}}\nc\cY{{\mathcal Y}}
\nc\bfz{{\boldsymbol z}}\nc\bfZ{{\boldsymbol Z}}\nc\cZ{{\mathcal Z}}

\newcommand{\remove}[1]{}

\newtheorem{theorem}{Theorem}
\newtheorem{definition}{Definition}
\newtheorem{lemma}[theorem]{Lemma}

\newcommand\reals{{\mathbb R}}

\newcommand\ff{{\mathbb F}}

\newcommand{\Om}{\Omega}
\newcommand{\eps}{\epsilon}
\newcommand{\nnz}{\mathrm{nnz}}

\newcommand{\mLabel}[1]{\mbox{$\scriptstyle{#1}$}}
\font\btt=rm-lmtk10

\begin{document}

\title{Low rank approximation and decomposition of large matrices using error
correcting codes}

\author{Shashanka Ubaru,
        Arya Mazumdar,
        and Yousef Saad
\thanks{S. Ubaru and Y. Saad are with the Department of Computer Science and
Engineering,
            University of Minnesota, Twin Cities, MN USA
            (e-mail: ubaru001@umn.edu, saad@cs.umn.edu).
A. Mazumdar is with Department of Electrical and Computer Engineering,
            University of Minnesota, Twin Cities, MN USA
            (e-mail: arya@umn.edu).
            This work was supported by NSF under grant NSF/CCF-1318597
(S. Ubaru and Y. Saad) and NSF/CCF-1318093 (A. Mazumdar).
A preliminary version of this work has appeared as conference proceedings 
in the 32nd International Conference on Machine Learning \cite{ubarulow}.
}}

\maketitle

\begin{abstract}
Low rank approximation is an important tool used in many applications of signal
processing and machine learning.
 Recently, randomized sketching algorithms were 
proposed to effectively construct low rank approximations and obtain approximate
singular value 
decompositions of large matrices. Similar ideas were used to solve least squares
regression problems.
In this paper, we show how matrices from error correcting codes can be used to
find such
low rank approximations  and matrix decompositions, and
 extend the framework to linear least squares  regression  problems.

The benefits of using these code matrices are the following: 
(i) They are easy to generate and they reduce randomness significantly. 
(ii) Code matrices, with mild restrictions, satisfy the subspace embedding property, 
and have a better chance of preserving 
the geometry of an large subspace of vectors.
(iii) For parallel and distributed applications,
code matrices have significant advantages over 
 structured random matrices and Gaussian random matrices.
(iv) Unlike Fourier or Hadamard transform matrices, which require sampling  
$O(k\log k)$ columns for a rank-$k$ approximation, 
the log factor is not necessary for certain types of code matrices.
In particular, $(1+\eps)$ optimal Frobenius norm error can be achieved for a rank-$k$
approximation with $O(k/\eps)$ samples.
(v) Fast multiplication is  possible 
with structured code matrices,
so fast approximations can be achieved for general dense input matrices.
(vi) For least squares regression problem $\min\|Ax-b\|_2$ where $A\in \reals^{n\times d}$,
 the $(1+\eps)$ relative error approximation can be achieved 
 with $O(d/\eps)$ samples,
with high probability, when certain code matrices are used.
\end{abstract}
\IEEEpeerreviewmaketitle 

\section{Introduction}

\IEEEPARstart{M}{any}  scientific computations, signal processing, data analysis 
and machine learning 
applications lead to  large dimensional matrices that can be
well approximated by 
a low dimensional (low rank) basis \cite{lra-book,lowrank1,review,drineas2006fast}. 
It is more efficient to solve many computational problems  by first transforming
these high dimensional 
matrices
into a low dimensional space, while preserving the invariant subspace that
captures the essential 
information of the matrix. 
Low-rank matrix approximation is an integral component of tools such as
principal 
component analysis (PCA) \cite{jolliffe2002principal}. It is also an
important instrument used in many applications 
like  computer vision (e.g., face recognition) \cite{turk1991eigenfaces}, signal
processing (e.g., adaptive beamforming) 
\cite{parker2005signal},
recommender systems \cite{drineas2002competitive},
information retrieval and latent
 semantic indexing 
 \cite{berry1995using,inforetrieval}, web search modeling \cite{kleinberg1999authoritative},
 DNA microarray data
 \cite{alter2000singular,raychaudhuri2000principal} and text mining, to name a few examples.
Several algorithms have been proposed in the literature for finding  low
rank approximations 
of  matrices
\cite{lra-book,lowrank1,review,drineas2006fast,boutsidis2009improved}.
Recently, research focussed  on developing techniques that use randomization for
computing low rank
approximations and  decompositions of such large matrices
\cite{review,sarlos2006improved,liberty2007randomized,
woolfe2008fast,nguyen2009fast,clarkson2009numerical,mahoney2011randomized,
woodruff2014sketching}.
It was found that randomness provides an effective way to construct low
dimensional bases with high 
reliability and computational efficiency.
Similar ideas based on random sampling have been proposed in the recent literature for solving 
least squares ($\ell_2$) linear regression problems
\cite{drineas2006sampling,sarlos2006improved,rokhlin2008fast,
clarkson2009numerical,nguyen2009fast,drineas2011faster,clarkson2013low}.   

Randomization techniques for matrix approximations  aim
to compute  
a basis that  approximately spans the range of an $m\times n$ input  matrix $A$,
by sampling\footnote{Sampling is sometimes called `sketching', popular in data streaming 
model applications \cite{clarkson2009numerical,woodruff2014sketching}.} 
the matrix $A$ using  random matrices, e.g. i.i.d Gaussian  \cite{review}. 
 This task is accomplished by first forming the matrix-matrix 
 product $Y=A\Omega$,  where $\Omega$ is an $n\times \ell$ random matrix of 
 smaller dimension $\ell\ll\{m,n\}$, 
 and then computing the orthonormal basis of $Y=QR$ that identifies the range of
the reduced matrix $Y$.
 It can be shown that  $A\approx QQ^\top A$ with high probability.
It has been shown that structured random matrices, 
like subsampled random Fourier transform (SRFT) and Hadamard transform (SRHT) 
matrices can also be used in place of   fully random matrices
\cite{woolfe2008fast,liberty2009accelerated,nguyen2009fast,tropp2011improved}. In this paper, we 
demonstrate how matrices from error correcting codes can be a good choice for 
computing such low rank approximations.

The input matrices, whose low rank approximation is to be computed, usually have
very 
large dimensions (e.g., in the order of $10^6-10^9$ \cite{review,yang2015implementing}).
In order to form a Gaussian (a fully) random  matrix that samples the input matrix, 
we need to  generate a large quantity of  random numbers. This could
be a serious 
practical issue
 (in terms of time complexity and storage). This issue can be addressed by using
the structured
 random matrices, like SRFT and SRHT matrices.
An important practical advantage of using  these structured
 random matrices is that their structure allows the computation of matrix-matrix product
at a cost of $O(mn \log_2\ell)$ making the algorithms fast (also known as fast transforms)
for general dense input matrices.
However, with these matrices,
mixing of columns might not be as uniform (some of the entries of the 
Fourier transform  might have very large magnitude), and there is potential loss in the
accuracy (i.e., the performance of  SRHT/SRFT matrices in place of 
Gaussian matrices with same number of samples $\ell$ (or even slightly higher) is worse). 

Another drawback with fast transforms is that 
for parallel and distributed applications, particularly when the input matrix 
is sparse and/or its columns are distributively stored, it is found that FFT-like algorithms are 
significantly slower due to communication issues or other machine related issues
(machines are optimized for matrix-vector operations)
\cite{yang2015implementing}. 
Similar issues arise when the input matrices are streamed 
\cite{review,clarkson2009numerical};
we give more details on these 
in sec.~\ref{sec:diff} and \ref{sec:choice}.
Also for a rank-$k$ approximation, these matrices require sampling $\ell= O(k\log
k)$ columns. 
Other practical issues arise such as: the Fourier Transform matrices require
handling complex 
numbers and the Hadamard matrices exist only for the sizes which are in powers
of $2$.
All these drawbacks can be overcome if the code matrices presented in this paper are
used for sampling the input matrices.

In digital communication, information is encoded (by adding redundancy) to
(predominantly binary)
vectors or codewords, that are then  transmitted 
over a noisy channel  \cite{cover2012elements}. 
These codewords are required to be 
far apart in terms of some distance metric for noise-resilience.
Coding schemes usually generate codewords that maintain a fixed minimum Hamming
distance between each other, hence they are widespread and act like random
vectors. We can define probability measures for  matrices formed by stacking up these codewords
(see section \ref{sec:ECC} for details).
In this paper, we explore the idea of using subsampled versions of these code matrices  as sampling 
(sketching) matrices in the 
randomized techniques for matrix approximations.

The idea of using code matrices for  such applications is not new in the literature.
A class of dual BCH code matrices were used in
\cite{ailon2009fast,liberty2009accelerated} as 
Fast Johnson-Lindenstrauss Transform (FJLT) to perform fast dimensionality reduction
of vectors.
Code matrices have also been used in applications of sparse signal recovery, such as 
compressed sensing
\cite{barg2015restricted} and group testing \cite{du2000combinatorial,mazumdar2012almost,ubarugroup}.
For matrix approximations, it is important to show that the sampling matrices used
can approximately preserve the geometry of an entire subspace of vectors, i.e., they satisfy the 
``subspace embedding'' property \cite{sarlos2006improved,woodruff2014sketching}.
 In section \ref{sec:preserve}, we show that the subsampled code matrices 
 with certain mild properties satisfy this subspace embedding property  with 
 high probability.
 Similar to Fourier and Hadamard sampling matrices, fast multiplication 
 is possible with code matrices from
certain class of codes due to their structure (see section \ref{sec:cost}
for details).
Hence, fast approximations can be achieved for general dense input matrices, since the
matrix-matrix product $A\Om$ can be computed in $O(mn \log_2\ell)$ cost with such code matrices.

 In addition, the shortcomings of SRFT/SRHT matrices in  
 parallel and distributed environments, and in data streaming models
 can be overcome by using code matrices (details in secs.~\ref{sec:cost},\ref{sec:choice}).
For certain  code matrices,
 the logarithmic factor in the number of samples is not required
 (see sec.~\ref{sec:logfactor} for an explanation). 
 This is a significant theoretical result that shows that
order optimality can be achieved 
in the number of samples required with partially random matrices.
Similar improvements were posed as an open problem in \cite{deshpande2006adaptive}
and in \cite{nguyen2009fast}. In the context of sparse approximations such improvements 
appear as main results in many places, see Table 1 of
\cite{berinde2008practical}.

A preliminary version of part of this paper has appeared in the conference proceedings 
of the 32nd International Conference on Machine Learning \cite{ubarulow}.
In this paper, we  improve the theoretical results obtained in \cite{ubarulow}. 
In particular, we show that `$(1+\eps)$' optimal Frobenius norm error can be achieved 
for low rank approximation using code matrices with a mild condition.
We also improve the spectral norm error bound 
obtained in \cite{ubarulow}. 
Furthermore, we  give several additional details that 
were omitted in the shorter version \cite{ubarulow} such as, details on the computational
cost, the choice of the code matrices for different scenarios 
(type of input matrices,
computational environments, etc.) and others.
We also extend the theory to show how code matrices can be used to
solve linear least squares ($\ell_2$) regression  problems (see below).

One of the key applications where the randomized approximation (or sketching) algorithms 
are  used is in  approximately solving 
 overdetermined least squares regression problem faster 
 \cite{drineas2006sampling,sarlos2006improved,drineas2011faster,clarkson2013low}. 
Here, we are given a matrix $A\in\reals^{n\times d}$ and a vector $b\in\reals^n$, with
$n\gg d$. The goal is to solve the least squares regression problem $\min_x\|Ax-b\|_2$
faster, (where $\|.\|_2$ is $\ell_2$ norm)
and output a vector $x'$ such that, with high probability,
\[
 \|Ax'-b\|_2\leq(1+\eps)\|A\hat{x}-b\|_2,
\]
where $\hat{x}$ is the $\ell_2$ minimizer given by the Moore-Penrose pseudo inverse of $A$,
i.e., $\hat{x}=A^{\dag}b$ \cite{golub2012matrix}.
For details on the applications where we encounter such extremely 
overdetermined linear system of equations, we refer to \cite{yang2015implementing}.
The idea of randomized approximation \cite{drineas2006sampling}
is to use a sampling (sketching) matrix to reduce the dimensions of $A$ and $b$, and then solve the smaller 
problem to obtain $x'$. In section \ref{sec:l2reg}, we show how the code matrices can be
used as the sampling matrix in such least squares regression problems.

\subsection*{Our Contribution}
In this paper, we advocate the use of error correcting coding matrices for randomized sampling 
of large matrices in low rank approximations and other applications, and show how specific classes
 of code matrices can be used in different computational environments to achieve the 
 best results possible (amongst the existing sampling matrices).
 
From the technical/theoretical point of view, for the different classes of code matrices, 
we show that the $(1+\eps)$ optimal Frobenius and spectral norm error bounds 
can be achieved with different sampling complexities  for the 
different classes of codes. To the best of our knowledge such results have not appeared in the 
literature before.
This comprehensive theoretical analysis results are established by combining various 
results that are previously developed in the literature and the properties of the code matrices.
Previous works of~\cite{ailon2009fast,liberty2009accelerated} consider the
dual BCH codes and the results developed there
show that certain classes of code matrices (dual BCH codes with
dual distance $>4$) satisfy the 
Johnson-Lindenstrauss Lemma~\cite{johnson1984extensions}. We combine these results with 
some of the other results developed in the literature to show that, such code matrices 
also satisfy an important subspace embedding property~\cite{sarlos2006improved},
that is
required to derive the $(1+\eps)$ optimal error bound results. The sampling complexity required for
these sampling matrices will be $O((k \log k)/\eps)$, which is similar (same order) to  
the sampling complexity required for SRFT/SRHT matrices.

The key theoretical result we present is that, 
the $(1+\eps)$ optimal (Frobenius and spectral norms) error bounds 
can be achieved with $O(k/\eps)$ samples when  certain classes of code matrices
(code matrices with dual distance $>k$) are used 
for sampling. 
To the best of our knowledge, such achievability results  with almost deterministic
matrices  that is  order optimal  with  the immediate  lower bound  of
$O(k)$, has not appeared in the literature before. 
We discuss how different classes of code matrices 
with desired properties (for sampling) can be generated and used in practice.
We also discusses the advantages of code matrices over other classes of sampling matrices in the 
parallel and distributed environments, and also in the data streaming models. With the sizes of the datasets increasing rapidly, 
we believe such advantages of code matrices become significantly important, making them more appealing 
 for a range of applications where such randomized sampling is used. The competitive  status  of code matrices are 
summarize  in  Table~\ref{table:table0}.
We also provide numerical experiments that demonstrate the performance of code matrices in practice.

\paragraph*{Outline} The organization of the rest of this paper is as follows: Section \ref{sec:prelim} gives
the notation and key definitions used, the problem set up and a brief introduction to
error correcting coding techniques.
Section \ref{sec:construct} discusses the construction of the subsampled code matrices and
the intuition behind the construction. The algorithm of the present paper is described 
in section \ref{sec:algo} and its computational cost is discussed in 
section \ref{sec:cost}.
Section \ref{sec:analysis} discusses the error 
analysis for the algorithm, by deriving bounds for the Frobenius norm error, spectral norm error and 
the singular values obtained from the algorithm. We show that $(1+\eps)$ relative Frobenius norm error 
approximation can be achieved with  code matrices. For this, the code matrices need to satisfy two key 
properties which are discussed in section \ref{sec:preserve}. The bounds for the approximation errors 
and the singular values obtained are derived in section \ref{sec:error}.
In section \ref{sec:l2reg}, we extend the framework
to linear least squares ($\ell_2$) regression  problem and in section \ref{sec:choice},
we discuss the choice of error correcting codes for different types of input matrices and 
computational environments.
Section \ref{sec:results} illustrates the performance of code matrices via few numerical
experiments.

\section{Preliminaries}\label{sec:prelim}
First, we present some of the notation used, some key definitions and give a brief description of
error correcting codes as will be required for our purpose. 
\subsection{Notation and Definitions}
Throughout the paper, $\|\cdot\|_2$ refers to the $\ell_2$ or spectral norm. We use
$\|\cdot\|_F$ for the Frobenius norm.
The singular value decomposition (SVD) of a matrix $A$ is denoted by 
$A=U\Sigma V^\top $ and the singular values by
$\sigma_j(A)$.
We use $e_j$ for the $j$th standard basis vector. 
Given a  subset $T$ of indices in $\{1,\ldots,2^r\}$ with size $n$ and
$r\geq\lceil\log_2n\rceil$, we define a restriction (sampling) operator
$S_T:\mathbb{R}^{2^r}\rightarrow\mathbb{R}^T$ to be the projection of a vectors on to the coordinates of $T$:
\[
 S_T\bfx=(x_j)_{j\in T} .
\]
A Rademacher random variable takes values $\pm1$ with equal probability.
We write $\varepsilon$ for a Rademacher variable.

Next, we define two key properties which will be used frequently in our theoretical analysis. 
In order to use the subsampled code matrices as sampling 
matrices, they need to satisfy these two properties, see section \ref{sec:analysis}. 
First is the Johnson-Lindenstrauss Transform (JLT) \cite{johnson1984extensions} 
which played a key role in 
the development of embedding-based randomized sampling. 
Sarlos \cite{sarlos2006improved} developed the important relation between JLT and 
random matrix sampling (also known as subspace embedding), see details in section 
\ref{sec:preserve}. The JLT property is defined as  \cite{sarlos2006improved}:
\begin{definition}[Johnson-Lindenstrauss Transform]
  A matrix $\Om\in \mathbb{R}^{n\times \ell}$ forms a Johnson-Lindenstrauss Transform with 
  parameters $\eps,\delta,d$ or 
  JLT($\epsilon,\delta,d$) for any $0<\eps,\delta<1$,
 if for any $d$-element set $V\subset \mathbb{R}^{n}$, and
  for all $v\in V$ it holds 
 \[
  (1-\epsilon)\|v\|_2^2\leq\|\Om^\top v\|_2^2\leq(1+\epsilon)\|v\|_2^2
 \]
 with probability $1-\delta$.
\end{definition}

The other key property which the code matrices need to satisfy is the subspace embedding 
property  defined below.
\begin{definition}[Subspace Embedding]
  A matrix $\Om\in \mathbb{R}^{n\times \ell}$ is a $(1\pm\eps)$ $\ell_2$-subspace embedding
  for the row space of an $m\times n$ matrix $A$, if for an orthonormal basis
  $V\in \mathbb{R}^{n\times k}$ that spans the row space of $A$,  for all
  $x\in \mathbb{R}^{k}$
 \[
  \|\Om^\top Vx\|_2^2=(1\pm\eps)\|Vx\|_2^2=(1\pm\eps)\|x\|_2^2,
 \]
  where $\|\Om^\top Vx\|_2^2=(1\pm\eps)\|x\|_2^2$ stands for 
$(1-\epsilon)\|x\|_2^2\leq\|\Om^\top Vx\|_2^2\leq(1+\epsilon)\|x\|_2^2$.
\end{definition}

The above definition is useful when the sampling is achieved column-wise. A similar 
definition for row-wise sampling holds for an orthonormal matrix $U\in \mathbb{R}^{m\times k}$
which spans the column space of $A$, see \cite{woodruff2014sketching}. 
The above definition simplifies to the following condition: 
 \begin{equation}\label{eq:SUP}
  \|V^\top \Om\Om^\top V-I\|_2\leq\eps.
 \end{equation}
 The matrix $\Om$ with subspace embedding property, satisfying the above condition 
 is  said to approximately preserve the geometry of an entire subspace
of vectors \cite{tropp2011improved}. 

In low rank approximation methods, we compute an orthonormal basis that
approximately spans the range of an $m\times n$
input matrix $A$. That is, a matrix $Q$ having orthonormal
columns such that 
$A\thickapprox QQ^\top A$.
The basis matrix $Q$ must contain as few columns as possible, but it
needs to be an accurate approximation of the input matrix. I.e., we seek a matrix
$Q$ with $k$ orthonormal columns such that   
\begin{equation}\label{eq:tol}
  \|A-QQ^\top A\|_{\xi}\leq e_k,
\end{equation}
for a positive error tolerance $e_k$ and  $\xi \in \{2,F\}$.

The best rank-$k$ approximation of $A$ with respect to both Frobenius and spectral norm 
 is given by the Eckart-Young theorem \cite{theomin}, and it is
 $A_k=U_k\Sigma_kV_k^\top $, where $U_k$ and $V_k$ are the $k$-dominant
left and right singular vectors of $A$, respectively and diagonal $\Sigma_k$ 
contains the top $k$ singular values  of $A$. So, the optimal 
$Q$ in \eqref{eq:tol} will be $U_k$ for $\xi\in\{2,F\}$, and $e_k=\sigma_{k+1}$ for $\xi=2$
and $ e_k=\sum^n_{j=k+1}\sigma_{j}$ for $\xi=F$.
In the low rank approximation applications, the rank $k$ 
will be typically much smaller than $n$, and it can be
 computed fast using the recently proposed fast numerical rank estimation 
methods~\cite{ubaru2016fast,ubaru2017fast}.

  \subsection{Error Correcting Codes}\label{sec:ECC}
   
   In communication systems, data are transmitted from a source (transmitter) to
a destination (receiver) through physical channels. 
   These channels are usually noisy, causing errors in the data received. In
order to facilitate  detection and correction of these errors in the
receiver, error correcting codes  are used \cite{macwilliams1977theory}.
   A block of information (data) symbols are encoded into a binary
vector\footnote{Here, and in the rest of the text, we
   are considering only binary codes. In practice, codes over other alphabets
are also quite common.}, also called a codeword.
   Error correcting coding methods check the correctness of the codeword
received.
 The set of codewords corresponding to  a set of data vectors (or symbols) that
can possibly be transmitted is called the {\em code}.
 As per our definition a code $\cC$ is a subset of the binary vector space  of dimension $\ell$, $\ff_2^\ell$,
 where $\ell$ is an
integer.
 
 A code is said to be linear  when adding two codewords of the code
 coordinate-wise using modulo-2 arithmetic  results in a third codeword
 of the code.  Usually a linear code $\cC$ is represented by the tuple
 $[\ell, r]$,  where $\ell$  represents the codeword  length and  $r =
 \log_2 |\cC|$ is  the number of information bits  that can be encoded
 by the code. There are  $\ell-r$ redundant bits in the codeword, which
 are sometimes called parity  check bits, generated from messages using
 an appropriate rule.  It is not  necessary for a codeword to have the
 corresponding  information bits as  $r$ of  its coordinates,  but the
 information must be uniquely recoverable from the codeword.

It is perhaps obvious that a linear code $\cC$ is a linear subspace of dimension
$r$ in the
vector space $\ff_2^\ell$. The basis of $\cC$ can be written as the rows of a
matrix, which is 
known as the generator matrix of the code. 
The size of the generator matrix $G$ is  $r\times \ell$, and for any information
vector $\bfm \in \ff_2^r$, the
corresponding codeword is found by the following linear map:
\[
\bfc = \bfm G.
\]
Note that  all the arithmetic operations above are over the binary field
$\ff_2$.

To encode $r$ bits, we must have  $2^r$ unique codewords. Then, we may form a
matrix of  size $2^r\times \ell$ by stacking up all codewords that are formed by
the generator matrix of a given linear coding scheme,
\begin{equation}\label{eq:codeword}
  \underbrace{C}_{2^r\times \ell}=\underbrace{M}_{2^r\times
r}\underbrace{G}_{r\times \ell}.
\end{equation}
For a given tuple $[\ell,r]$, different  error correcting coding schemes
 have different generator matrices and the resulting  codes have  different
properties. 
For example, for any two integers $t$ and $q$, a  BCH code \cite{bose1960class}
has length $\ell = 2^q -1$ and
dimension $r = 2^q-1-tq$. Any two codewords in this BCH code maintain
 a   minimum (Hamming) distance of at least $2t+1$ 
between them. The  minimum pairwise distance is an important parameter of a code
and is called just the  distance of the code.

As a linear code $\cC$ is a subspace of a vector space, the null space $\cC^{\bot}$ of 
the code is
another well defined
subspace. This is called the {\em dual} of the code. For example, the dual of the $[2^q-1,
2^q -1-tq]$-BCH code
is a code with length $2^q-1$, dimension $tq$ and minimum distance at least
$2^{q-1} - (t-1)2^{q/2}.$
The minimum distance of the
dual code is called the  {\em dual distance} of the code.

Depending on the coding schemes used, the  codeword matrix $C$ will  
have a variety of favorable properties, e.g., low coherence which is 
useful in compressed sensing \cite{mazumdar2011combinatorial,barg2015restricted}. 
Since the codewords need to be  far apart, they
show some properties of random vectors. We  can define probability 
 measures for codes generated from a given coding scheme.
 If $\cC\subset\{0,1\}^{\ell}$ is an $\ff_2$-linear code whose dual  $\cC^{\bot}$
 has a minimum distance above $k$ (dual distance $> k$), then 
 the code matrix is an {\em orthogonal array} of strength $k$ \cite{delsarte1998association}.
%
This means, in  such a code $\cC$,  for any $k$ entries of a randomly and uniformly chosen
codeword $\bfc$ say 
 $\mathbf{c'}=\{c_{i_1},c_{i_2},\ldots,c_{i_k}\}$ and for any  
 $k$ bit binary string $\alpha$, we have
 \[
  \mathbf{Pr}[\mathbf{c'}=\alpha]=2^{-k}.
 \]
 This is called the $k$-wise independence property of codes.
We will use this  property of codes in our theoretical analysis
(see section \ref{sec:analysis} for details).

The codeword matrix $C$ has  $2^r$ codewords each of length $\ell$ (a $2^r\times
\ell$ matrix), 
i.e., a set of $2^r$ vectors in $\{0,1\}^{\ell}$.  
Given a codeword $\mathbf{c}\in \cC$, let us map it to a vector
$\phi\in \mathbb{R}^{\ell}$ by setting 
$1\longrightarrow \frac{-1}{\sqrt{2^{r}}}$ and
$0\longrightarrow\frac{1}{\sqrt{2^{r}}}$. 
In this way, a binary code $\cC$ gives rise to a code matrix
$\Phi=(\phi_1^\top,\ldots,\phi_{2^r}^\top)^\top $. 
Such a mapping  is called binary phase-shift keying (BPSK) and  appeared in the
context of sparse recovery (e.g., p.~66
 \cite{mazumdar2011combinatorial}). 
For  codes with dual distance $\geq3$, this code matrix $\Phi$ will have
orthonormal columns, see lemma \ref{lemm:orth}.
In section \ref{sec:preserve}, 
we will show that these code matrices with certain mild properties satisfy the JLT and the 
subspace embedding properties and
preserve the geometry of  vector subspaces with high probability.
In the randomized techniques for matrix approximations, we can use a subsampled
and scaled version of this  matrix
$\Phi$  to sample a given input matrix and  find its active subspace.

 \section{Construction of Subsampled Code Matrix}\label{sec:construct}
For an input matrix $A$ of size $m\times n$ and a target rank $k$, we choose
$r\geq\lceil\log_2 n\rceil$ as the dimension of the code (length of the message vector)
and $\ell>k$ as the length of the code. The value of $\ell$ will depend
 on the coding scheme used, particularly on the dual distance of the code 
 (details in section \ref{sec:preserve}).
 We consider an $[\ell,r]$-linear coding scheme and form the sampling matrix as
follows:
 We draw the sampling test matrix say $\Omega$ as
\begin{equation}\label{eq:SCM}
 \Omega=\sqrt{\frac{2^r}{\ell}}DS\Phi,
\end{equation}
where
\begin{itemize}
 \item $D$ is a random $n\times n$ diagonal matrix whose entries are independent
random signs, i.e., random variables 
 uniformly distributed on $\{\pm1\}$.
 \item $S$ is a uniformly random downsampler, an $n\times 2^r$matrix whose $n$
rows are randomly selected from a $2^r\times2^r$ identity matrix.
 \item $\Phi$ is the $2^r\times \ell$ code matrix, generated using an
$[\ell,r]$-linear coding scheme, with BPSK mapping and scaled by $2^{-r/2}$ such
that  all columns have unit norm.
\end{itemize}
 
\subsection{Intuition}
The design of a Subsampled Code Matrix (SCM) is similar to the design of SRFT and SRHT
matrices.
The intuition for using such a design is well established in 
\cite{tropp2011improved,review}.
The matrix $\Phi$ has entries with magnitude $\pm2^{-r/2}$ and has orthonormal
columns
when a coding scheme with dual distance of the code $\geq3$  is used.

 The  scaling $\sqrt{\frac{2^r}{\ell}}$ is used to make the energy of the
sampling matrix equal to unity, i.e., to make the rows of $\Omega$ unit
vectors. The objective of multiplying by the matrix $D$ is twofold.
 The first purpose is to flatten out the magnitudes of input vectors, see
\cite{tropp2011improved} for the  details.
For a fixed unit vector $\bfx$,  the first component of $\bfx^\top DS\Phi$ is given
by
 $(\bfx^\top DS\Phi)_1=\sum_{i=1}^nx_i\varepsilon_i\phi_{j1}$,
where $\phi_{j1}$ are components of the first column of the code matrix $\Phi$, with the indices $j$'s
are such that $S_{ij}=1$ for $i=1,\ldots,n$ and $\varepsilon_i$ is the Rademacher variable
from $D$.
This sum  has zero mean and since entries of $\Phi$ have magnitude
$2^{-r/2}$, the variance of the sum is $2^{-r}$.
The Hoeffding inequality \cite{hoeffding1963probability} shows that 
\[
 \mathbb{P}\{|(\bfx^\top DS\Phi)_1|\geq \tilde{t}\}\leq 2e^{-2^r\tilde{t}^2/2}.
\]
That is, the magnitude of the first component of $\bfx^\top DS\Phi$ is about
$2^{-r/2}$. Similarly, the argument holds for the remaining entries. 
Therefore, it is unlikely that any one of the $\ell$ components of 
$\bfx^\top DS\Phi$ is larger than $\sqrt{4\log \ell/2^r}$
(with a failure probability of  $2\ell^{-1}$). 

The second purpose of multiplying by $D$ is as follows:
The code matrix $\Phi$ with a dual distance $>k$ forms a deterministic $k$-wise
independent matrix. Multiplying this $\Phi$ matrix  by
$D$ (with independent random signs on the diagonal) results in a $k$-wise independent random matrix.
Note that uniform downsampling of the matrix will not affect this
property. Hence, the subsampled code matrix SCM $\Om$ will be a $k$-wise independent random matrix.
This is a key property of SCM $\Om$ that we will use to prove the JLT and the subspace 
embedding properties for SCM, see section \ref{sec:preserve}.

The downsampler $S$ is a formal way of saying, if $n<2^r$, we choose $n$ out of $2^r$
possible codewords to form the sampling matrix $\Om$.
Uniform downsampling is used in the theoretical
analysis to get an upper bound for the singular values of $\Om$ (see sec.~\ref{sec:error}).
In practice, we choose $n$ numbers between $1$ to $2^r$, use the binary representation of these 
numbers as the message vectors (form $M$) and use the generator matrix $G$ of the coding scheme
selected to form the sampling matrix $\Om$, using \eqref{eq:codeword} and BPSK mapping.
For dense input matrices, it is advantageous to choose these numbers (message vectors) to be
$1$ to $2^{\lceil\log_2 n\rceil}$, to exploit the availability of fast multiplication (see
details in section \ref{sec:cost}).
\section{Algorithm}\label{sec:algo}
 We use the same prototype algorithm as 
discussed in \cite{review} for the low rank approximation and decomposition of an
input matrix $A$.
The subsampled code matrix (SCM) $\Om$ given in (\ref{eq:SCM}), generated from a
chosen coding scheme 
is used as the sampling test matrix. The algorithm is as follows:
\begin{algorithm}[H]
\caption{Prototype Algorithm}
\label{alg:algo1}
\begin{algorithmic}
   \STATE {\bfseries Input:} An $m\times n$ matrix $A$, a target rank $k$.
 \STATE {\bfseries Output:} Rank-$k$ factors $U,\Sigma$, and $V$ in an
approximate
SVD $A\thickapprox U\Sigma V^\top $.
\STATE {\bfseries 1.} Form an $n\times \ell$ subsampled code matrix $\Omega$, as
described in
Section \ref{sec:construct} and  (\ref{eq:SCM}), using an $[\ell,r]-$linear
coding scheme, 
where $\ell>k$ and $r\geq\lceil\log_2 n\rceil$. 
\STATE {\bfseries 2.} Form the $m\times \ell$ sample matrix $Y=A\Omega$.
\STATE {\bfseries 3.} Form an $m\times \ell$ orthonormal matrix $Q$ such that\\ 
$Y=QR$. 
\STATE {\bfseries 4.}  Form the $\ell\times n$ matrix $B=Q^\top A$.
\STATE {\bfseries 5.}  Compute the SVD of the small matrix $B=\hat{U}\Sigma
V^\top $.
\STATE {\bfseries 6.} Form the matrix $U=Q\hat{U}$.
\end{algorithmic}
\end{algorithm}

The prototype algorithm requires only two passes over the input matrix 
(single pass algorithms can also be developed \cite[\S 5.5]{review}),
as opposed to $O(k)$ passes required 
for classical algorithms. This is particularly significant when the input
matrix is very large 
to fit in fast memory (RAM) or when the matrices are streamed \cite{review}.
It is known that, the randomized techniques allow us to reorganize the 
calculations required to exploit the input 
matrix properties and the modern computer architecture more efficiently. 
The algorithm is also well suited for implementation in parallel and
distributed environments, see \cite{yang2015implementing}. For more 
details on all the advantages
of randomized methods over classical techniques, we refer to
\cite{sarlos2006improved,woolfe2008fast,review}.

Several algorithms have been developed in the literature 
which  build on the above prototype algorithm.
An important requirement (rather a drawback) of the prototype algorithm is that, 
to obtain a good approximation, the algorithm requires the
 singular values of the input matrix to 
 decay rapidly \cite{review}. Methods such as 
randomized power method \cite{review,rokhlin2009randomized,mingu2014} and  
randomized block Krylov subspace methods \cite{musco2015stronger}
have been proposed to  improve the performance (accuracy) of the  prototype algorithm,
particularly when the singular values of the input matrix decay slowly.
In these methods, step 2 in Algorithm~\ref{alg:algo1} 
is replaced by $Y=(AA^\top )^{q'}A\Omega$, where $q'$
is a small integer, or a block Krylov subspace \cite{golub2012matrix}. 
However, these algorithms require  $2(q'+1)$ passes over $A$.
Use of structured random matrices like SRFT and SRHT are proposed for a faster  computation of 
the matrix product $Y=A\Omega$ \cite{woolfe2008fast,liberty2009accelerated},
for dense input matrices.
The use of sparse random matrices, e.g. \texttt{CountSketch} matrix
\cite{clarkson2013low,woodruff2014sketching}, is
 proposed to achieve faster computations when the input matrix is sparse.
 
Algorithm \ref{alg:algo1} can also be modified to obtain the eigenvalue decompositions of 
 square input matrices \cite{review}.
In all the above mentioned modified algorithms, we can use our subsampled code matrix
as the random sampling (sketching) matrix.
For the analysis in the following sections, 
we shall consider the prototype algorithm \ref{alg:algo1}.

\section{Computational Cost}\label{sec:cost}
One of the key advantages of using structured random matrices (SRFT or SRHT) 
in the randomized sketching algorithms 
is that, for a general dense matrix, we can compute the matrix-matrix product $Y=A\Omega$
in $O(mn\log_2 \ell)$ time exploiting the structure of Fourier/Hadamard matrices
\cite{sarlos2006improved,woolfe2008fast,nguyen2009fast,liberty2009accelerated,review}.
The idea of fast multiplications was inspired by articles on 
Fast Johnson-Lindenstrauss Transform (FJLT) \cite{ailon2006approximate,ailon2009fast} 
where it was shown that matrix-vector products with such structured matrices 
can be computed in $O(n\log_2\ell)$ time.
Interestingly, Ailon and Liberty \cite{ailon2009fast} give dual BCH code matrices 
and Hadamard matrices (that are actually a special
codes called 
1st order Reed-Muller codes) as examples for such structured matrices.

Many, if not  most of the structured  codes can be decoded using the
Fast Fourier  Transform (FFT) \cite{blahut1979transform}.
The corresponding $2^r\times \ell$ code matrix $\Phi$ of such structured codes 
(after BPSK mapping) will have 
every column of $\Phi$ equal to some column of a $2^r\times 2^r$ Hadamard matrix,
see definition 2.2 in \cite{ailon2009fast}. Hence, 
for a general dense matrix in RAM,
the  matrix-matrix product $Y=A\Omega$
with these structure code matrices can be computed in $O(mn\log_2 \ell)$ time  
using the `Trimmed Hadamard transform' technique
described in \cite{ailon2009fast,liberty2009accelerated}.
If $n<2^r$, we choose the top $2^{\lceil\log_2 n\rceil}$ codewords of $\Phi$ as the rows of 
$\Om$ such that the
columns of $\Om$ are some columns of a $2^{\lceil\log_2 n\rceil}\times2^{\lceil\log_2 n\rceil}$
Hadamard matrix.

Fast multiplications are possible with  matrices from another 
class of codes known as
cyclic codes. In cyclic codes, a circular shift of a codeword results in another codeword of 
that code. So, a $2^r\times \ell$ code matrix $\Phi$ generated using an
$[\ell,r]$-cyclic code scheme will consist of $2^r/\ell$ blocks of circulant matrices
of size $\ell\times\ell$ (when appropriately rearranged). It is known that 
the matrix-vector products with  circulant matrices 
can be computed in $O(\ell\log_2 \ell)$ operations via FFT \cite{golub2012matrix}.
So, for a general dense input matrix, the  matrix-matrix product $Y=A\Omega$
with such cyclic code matrices can be computed in $O(mn\log_2 \ell)$ time.

The remaining steps (steps $3-6$) of the algorithm can be computed in $O((m+n)k^2)$
time using the row extraction method described in \cite{review}. Therefore, 
for a general dense input  matrix in RAM,
the total computational cost of Algorithm \ref{alg:algo1}
using SCM is $O(mn\log_2\ell+(m+n)k^2)$ for structured and cyclic codes.

For sparse input matrices or when the columns of $A$ are distributively stored, 
we can choose codewords at random from a desired code (as described earlier)
making $\Om$ unstructured and $Y=A\Om$ a dense transform, similar to a 
random sampling matrix. 
The computational cost of the algorithm  for such cases is $O(\nnz(A)\ell+(m+n)k^2)$, 
where $\nnz(A)$ is the number of 
nonzero entries in the input matrix $A$.
We will see that, for code matrices with  certain properties, $\ell=O(k/\eps)$ 
which will be advantageous in these cases (compared to SRFT/SRHT which require 
$\ell=O(k\log k/\eps)$).
Additional details of the choice of the code matrix for different types of input matrices and 
 computational environments are given in section \ref{sec:choice}.

\section{Analysis}\label{sec:analysis}
This section discusses the performance (error) analysis of the subsampled code matrices (SCM)
as sampling matrices 
in Algorithm \ref{alg:algo1}. We will prove that an approximation error of $(1+\eps)$ times
the best rank-$k$ approximation (Frobenius norm error) possible for a given matrix $A$
can be achieved with code matrices. That is,
\[
 \|A-\hat{A}_k\|_F\leq(1+\eps)\|A-A_k\|_F,
\]
where $\hat{A}_k$ is the rank-$k$ approximation obtained from  Algorithm \ref{alg:algo1} 
and $A_k$ is the best rank-$k$ approximation as defined in section \ref{sec:prelim}.
In order to prove  this, we show that SCM satisfies the Johnson Lindenstrauss Transforms (JLT) and
the subspace embedding properties via 
the $k$-wise independence property of the codes (the relation between these two properties and 
$(1+\eps)$ approximation is given in sec.~\ref{sec:error}, Lemma \ref{lemm:2cond}).
We also derive the bounds for the spectral norm error and the singular
values obtained, based on the deterministic error bounds in the literature
for the algorithm for a given sampling matrix $\Om$.
\subsection{Setup} 
 Let $A$ be an $m\times n$ input matrix with SVD given by
 $A=U\Sigma V^\top$,
and   partition its SVD as follows
 \begin{equation}\label{eq:partition}
  A=U\begin{blockarray}{c@{}cc@{\hspace{5pt}}cl}
    & \mLabel{k} & \mLabel{n-k}& & \\
    \begin{block}{[c@{\hspace{5pt}}cc@{\hspace{5pt}}c]l}
      &\Sigma_1 & & & \\
      & & \Sigma_2& & \\
    \end{block}
  \end{blockarray}
 \begin{blockarray}{c@{}c@{\hspace{5pt}}cl}
    & \mLabel{n} & & \\
    \begin{block}{[c@{\hspace{5pt}}c@{\hspace{5pt}}c]l}
      & V_1^\top  & & \mLabel{k} \\
      & V_2^\top  & & \mLabel{n-k} \\
    \end{block}
  \end{blockarray}.
\end{equation}
Let $\Omega$  be the $n\times \ell$ test (sampling) matrix, where $\ell$ is the
number of samples.
Consider the matrices
\begin{eqnarray}\label{eq:ompart}
 \Omega_1=V_1^\top \Omega&\text{and}& \Omega_2=V_2^\top \Omega.
\end{eqnarray}
The objective of any low rank approximation algorithm is to approximate
the subspace that spans the top $k$ left singular vectors of $A$.
 Hence, for a given sampling matrix $\Om$, the key challenge is to show  that 
$\Om_1$ is full rank. That is, we need to  show that for any orthonormal matrix  $V$ of dimension $k$, 
with high probability $V^\top \Omega$ is well conditioned \cite{review}. This is true if the 
 test matrix $\Omega$ satisfies the subspace embedding property, and it 
 is said to preserve the geometry of an entire subspace
of vectors $V$.

\subsection{Subsampled Code Matrices, JLT and Subspace Embedding}\label{sec:preserve}
Recall from section \ref{sec:construct} the construction of the `tall and thin'
$n\times \ell$ subsampled
error correcting code matrix $\Omega$.
The critical requirement to prove the $(1+\eps)$ optimal error bound
is to show that  these matrices satisfy the two key properties: JLT and subspace embedding. 
The subspace embedding property will also imply that $\Om_1$ will be full rank, which will enable 
us use the deterministic bounds developed in the literature to derive the bounds for 
the spectral norm error and the singular values obtained (see sec.~\ref{sec:deter}).
 \subsubsection{Johnson-Lindenstrauss Transform}
 We saw the definition of JLT in sec.~\ref{sec:prelim}, which says that a matrix $\Om$ that satisfies
  JLT($\epsilon,\delta,d$) preserves the norm for any vector $v$ in a $d$-element
  subspace $V\subset \mathbb{R}^{n}$.
We will use two key results developed in the literature to
show that code matrices with certain mild properties satisfy 
the JLT property. 

The first result is by Ailon and Liberty 
\cite{ailon2009fast}, where they  show a matrix
$\Om$ which is 4-wise independent will satisfy the JLT property, see Lemma~5.1 in 
\cite{ailon2009fast}. Interestingly, they give the
 2 error correcting dual BCH codes as examples for such 4-wise independent matrices
 and also demonstrate  how fast multiplications can be 
achieved with these code matrices. 
However, a minor drawback with using 4-wise independent matrices
is that the maximum entries of $A$ need to be restricted.

The second (stronger) result is by Clarkson and Woodruff 
\cite{clarkson2009numerical} (see Theorem 2.2), where they  show
if $\Om$  is a $4\lceil\log(\sqrt(2)/\delta)\rceil$-wise 
independent matrix, then $\Om$ will satisfy the JLT property. 
Recall that the SCM matrix $\Om$ defined in eq.~\eqref{eq:SCM} will be a random $k$-wise 
independent matrix if the dual distance of the code is $>k$.
Thus, any error correcting code matrix with a dual distance $>4$ 
(more than 2 error correcting ability) will  
satisfy the JLT property.

One of the important results related to JLT that is of interest for our theoretical analysis is the
matrix multiplication property. This 
is defined in the following lemma, 
which is Theorem 2.8 in \cite{woodruff2014sketching}. We can see similar results in 
Lemma 6 in \cite{sarlos2006improved} and Theorem 2.2 in \cite{clarkson2009numerical}.
\begin{lemma}\label{lemm:JLT}
 For $\eps,\delta\in (0,1/2)$, let $\Om$ be a random matrix (or from a distribution $\mathcal{D}$)
 with $n$ rows that satisfies $(\eps,\delta,d)$-JLT property. Then for $A,B$ matrices with $n$ 
 rows, 
 \begin{equation}
   \mathbf{Pr}\left[\|A^\top B-A^\top\Om\Om^\top B\|_F\leq 3\eps \|A\|_F  \|B\|_F\right]\geq1-\delta.
 \end{equation}
\end{lemma}
We will see in sec.~\ref{sec:error} that the above
lemma is one of the two main ingredients required to prove  $(1+\eps)$ optimal error bound.
The other ingredient is the subspace embedding property.

\subsubsection{Subspace Embedding}
One of the primary results developed in the randomized sampling algorithms literature 
 was establishing the relation between the Johnson-Lindenstrauss Transform (JLT) and subspace
 embedding.
The following lemma which is  corollary 11 in  \cite{sarlos2006improved}
gives this important relation.
\begin{lemma}\label{lemm:lemma1}
Let $0<\eps,\delta<1$ and $f$ be some function.
If $\Om\in \mathbb{R}^{n\times \ell}$ satisfies a JLT-$(\eps,\delta,k)$ with 
$\ell=O(k\log(k/\eps)/\eps^2.f(\delta))$, then for any orthonormal matrix 
$V\in\mathbb{R}^{n\times k},n\geq k$ we have
\[
 \mathbf{Pr}( \|V^\top \Om\Om^\top V-I\|_2\leq\eps)\geq1-\delta. 
\]
\end{lemma}

The above lemma shows that, any sampling matrix $\Om$ satisfying JLT and having length 
$\ell=O(k\log(k/\eps)/\eps^2)$ satisfies the subspace embedding property.
Thus, any SCM $\Om$  with a dual distance $>4$ will also satisfy the subspace embedding property
(since they satisfy JLT as we saw in the previous section).
The subspace embedding property implies that the 
 singular values of $V^\top \Om$ are bounded, i.e., $V^\top \Om$ is well conditioned 
with high probability. This result is critical since it shows that the SCM matrices can preserve the 
geometry of the top $k$-singular vectors of the input matrix $A$.

Observe that with the above analysis, we will require  $\ell=O(k\log(k/\eps))$
 number of samples  for the subspace embedding property to be satisfied,
which is similar to a subsampled 
Fourier or Hadamard matrix. Next, we show that for the subspace embedding property to be satisfied,
we will require only $O(k/\eps)$  number of samples
for certain types of code matrices. 

We know that the code matrices display some of the properties of 
 random matrices, particularly when the distance of the code is high.
 Indeed a code with dual distance above $k$ supports $k$-wise independent probability measure
 and SCM $\Om$ will be a random matrix with $k$-wise independent rows.
This property of SCM helps us  use 
the following lemma given in \cite[Lemma 3.4]{clarkson2009numerical} which states,
 \begin{lemma}
 Given an integer $k$ and $\eps,\delta>0$.
If $\Om\in \mathbb{R}^{n\times \ell}$ is 
 $\rho(k+\log(1/\delta))$-wise independent matrix with an absolute constant 
 $\rho>1$, then for any orthonormal matrix $V\in\mathbb{R}^{n\times k}$
 and $\ell=O(k\log(1/\delta)/\eps)$, with probability at least $1-\delta$ we have 
 \[
  \|V^\top \Om\Om^\top V-I\|_2\leq\eps.
 \]
 \end{lemma}
Thus, a sampling SCM matrix $\Om$ which is $\lceil k+\log(1/\delta)\rceil$-wise independent
satisfies  the subspace embedding property with the number of samples (length) $\ell=O(k/\eps)$.
Hence, an SCM  $\Om$  with dual distance $>\lceil k+\log(1/\delta)\rceil$
 will preserve the geometry of $V$ with $\ell=O(k/\eps)$.

In summary, any SCM with dual distance $>4$ satisfies the JLT property, and 
will satisfy the subspace embedding property
 if $\ell=O(k\log(k/\eps))$. If the dual distance is $>k$, then the SCM
 can preserve the geometry of $V$ with $\ell=O(k/\eps)$.
 \subsection{Deterministic Error bounds}\label{sec:deter}
 In order to derive the bounds for the spectral norm error and the singular values obtained, we will 
 use the deterministic error bounds for Algorithm \ref{alg:algo1} developed in the literature
 \cite{review,mingu2014}.
Algorithm \ref{alg:algo1} constructs an orthonormal basis $Q$ for the range of
$Y$, and the goal is to quantify
how well this basis captures the action of the input matrix $A$.
Let $QQ^\top =P_Y$, where $P_Y$ is the unique orthogonal projector with 
range($P_Y$)~=~range($Y$). If $Y$ is full rank, we can express the projector as :
$P_Y=Y(Y^\top Y)^{-1}Y^\top $.
We seek to find an upper bound for the approximation error given by,  for $\xi\in\{2,F\}$
\[
 \|A-QQ^\top A\|_{\xi}=\|(I-P_Y)A\|_{\xi}.
\]
The deterministic upper bound for the approximation error of Algorithm
\ref{alg:algo1} is given in \cite{review}.
 We restate theorem 9.1 in \cite{review} below:

\begin{theorem}[Deterministic error bound]\label{theo:1}
Let $A$ be $m\times n$ matrix with singular value decomposition given by
 $A=U\Sigma V^\top $, and fixed $k\geq0$. Choose a test matrix $\Omega$ and construct
the sample matrix $Y=A\Omega$. 
 Partition $\Sigma$ as in (\ref{eq:partition}), and define 
 $ \Omega_1$ and $ \Omega_2$ via (\ref{eq:ompart}). Assuming that $ \Omega_1$ is
full row rank, the approximation
 error satisfies for $\xi\in\{2,F\}$
 \begin{equation}\label{eq:approxerror}
  \|(I-P_Y)A\|_{\xi}^2\leq\|\Sigma_2\|_{\xi}^2+\|\Sigma_2 \Omega_2\Omega_1^{\dag}\|_{\xi}^2.
 \end{equation}
\end{theorem}
An elaborate proof for the above theorem can be found in \cite{review}.
Using the submultiplicative property of the spectral and Frobenius norms, 
and the Eckart-Young theorem mentioned earlier,
 equation \eqref{eq:approxerror} can be
simplified to
\begin{equation}\label{eq:errorbound}
 \|A-QQ^\top A\|_{\xi}\leq\|A-A_k\|_{\xi}\sqrt{1+\|\Omega_2\|_2^2\|\Omega_1^{\dag}\|_2^2}.
\end{equation}
Recently, Ming Gu \cite{mingu2014} developed deterministic lower bounds for the
singular values obtained
from randomized algorithms,
particularly for the power method \cite{review}.
Given below is the modified version of Theorem 4.3 in \cite{mingu2014} for
Algorithm \ref{alg:algo1}.

\begin{theorem}[Deterministic singular value bounds]\label{theo:2}
 Let  $A=U\Sigma V^\top $ be the SVD of $A$, for a fixed $k$, and let $V^\top \Omega$ be
partitioned as in (\ref{eq:ompart}).
Assuming that $ \Omega_1$ is full row rank, then Algorithm \ref{alg:algo1} must
satisfy for $j=1,\ldots,k$:
\begin{equation}\label{eq:errorbound2}
 \sigma_j\geq\sigma_j(\hat{A}_k)\geq\frac{\sigma_j}{\sqrt{1+\|\Omega_2\|_2^2\|\Omega_1^{
\dag}\|_2^2
 \left(\frac{\sigma_{k+1}}{\sigma_j}\right)^2}}
\end{equation}
where $ \sigma_j$ are the $j$th singular value of $A$ and $\hat{A}_k$ is the rank-$k$
approximation obtained by our algorithm. 
\end{theorem}
The proof for the above theorem can be seen in \cite{mingu2014}. In both the above theorems, the key 
assumption is that $ \Omega_1$ is full row rank. This is indeed true if the sampling matrix $\Om$
satisfies the subspace embedding property.

\subsection{Error Bounds}\label{sec:error}
The following theorem gives the approximation error bounds when the subsampled code matrix (SCM)
is used as the sampling matrix $\Omega$ in Algorithm \ref{alg:algo1}. 
The upper and lower bounds for the singular values obtained by the algorithm are also given.

\begin{theorem}[Error bounds for code matrix]\label{theo:BCH}
Let $A$ be $m\times n$ matrix with singular values
$\sigma_1\geq\sigma_2\geq\sigma_3\geq\ldots$. 
Generate a  subsampled code matrix $\Omega$ from a desired coding scheme as in
(\ref{eq:SCM}) with
$r\geq\lceil\log_2(n)\rceil$ as the dimension of the code. 
For any code matrix $\Om$ with {\bf dual distance $>4$ and length
$\ell=O(k\log(k/\eps)/\eps^2.f(\delta))$} the following  three bounds hold
with  probability at least $1-\delta$ :
\begin{enumerate}
 \item The Frobenius norm  error satisfies,
 \begin{equation}\label{eq:error1}
 \|A-\hat{A}_k\|_F\leq\|A-A_k\|_F(1+\eps).
\end{equation}
\item The spectral norm error satisfies,
\begin{equation}\label{eq:error2}
 \|A-\hat{A}_k\|_{2}\leq\|A-A_k\|_{2}\sqrt{1+\frac{3n}{\ell}}.
\end{equation}
\item The singular values obtained satisfy:
\begin{equation}\label{eq:singerror}
\sigma_j\geq\sigma_j(\hat{A}_k)\geq\frac{\sigma_j}{\sqrt{1+
\left(\frac{3n}{\ell}\right)
  \left(\frac{\sigma_{k+1}}{\sigma_j}\right)^2}}.
\end{equation}
\end{enumerate}
If the code matrix $\Om$ has {\bf dual distance $\geq \lceil k+\log(1/\delta)\rceil$}, 
then the above  three bounds hold for {\bf length
$\ell=O(k\log(1/\delta)/\eps)$}.
\end{theorem}
\begin{IEEEproof}[Proof - Frobenius norm Error]
As we have been alluding to in the previous sections, the  $(1+\eps)$ optimal Frobenius norm error given in
eq.~\eqref{eq:error1} is related to the JLT and the subspace embedding properties. 
The following lemma gives this relation which is Lemma 4.2 in  Woodruff's monograph \cite{woodruff2014sketching}.
\begin{lemma}\label{lemm:2cond} 
 Let $\Om$ satisfy the subspace embedding property for any fixed $k$-dimensional subspace
 $M$ with probability $9/10$, so that $\|\Om^\top y\|_2^2=(1\pm1/3)\|y\|_2^2$ for all $y\in M$.
 Further, suppose $\Om$ satisfies the $(\sqrt{\eps/k},9/10,k)$-JLT property such that the 
 conclusion in Lemma \ref{lemm:JLT} holds, i.e., for any matrices $A,B$ each with $n$ rows,
 \[
   \mathbf{Pr}\left[\|A^\top B-A^\top\Om\Om^\top B\|_F\leq 3\sqrt{\eps/k} \|A\|_F  \|A\|_F\right]\geq9/10.
 \]
 Then the column space of $A\Om$ contains a  $(1+\eps)$ rank-$k$ approximation to $A$.
\end{lemma}

From the analysis in section \ref{sec:preserve} (in particular from Lemma \ref{lemm:JLT} and 
\ref{lemm:lemma1}), we know that both the conditions in the above lemma 
are true for SCM  with dual distance $>4$ and length
$\ell=O(k\log(k/\eps)/\eps^2.f(\delta))$, when appropriate $\eps$ and $\delta$ are chosen. Since 
$\hat{A}_k=QQ^\top A$, where $Q$ is the orthonormal matrix spanning the column space of $A\Om$, we
obtain the Frobenius error bound in eq.~\eqref{eq:error1} from the above lemma.

Clarkson and Woodruff \cite{clarkson2009numerical} gave the Frobenius norm error bound 
for low rank approximation using $k$-wise independent sampling matrices. The error bound in
\eqref{eq:error1} for SCM with dual distance $>k$
is straight from the following lemma which is a modification of 
Theorem 4.2 in \cite{clarkson2009numerical}.
\begin{lemma}
 If $\Om\in\reals^{n\times\ell}$ is a $\rho(k+\log(1/\delta))$-wise independent sampling matrix,
 then for  $\ell=O(k\log(1/\delta)/\eps)$, with probability at least $1-\delta$, we have
 \begin{equation}
   \|A-\hat{A}_k\|_F\leq\|A-A_k\|_F(1+\eps).
 \end{equation}
\end{lemma}
Proof of this lemma is clear from the proof of Theorem 4.2 in  \cite{clarkson2009numerical}.
\end{IEEEproof}

\begin{IEEEproof}[Proof - Spectral norm Error]
The proof of  the approximate error bounds given in \eqref{eq:error2} follows 
from the deterministic bounds given in sec.~\ref{sec:deter}.
We start  from equation
\eqref{eq:errorbound} in Theorem \ref{theo:1},
the  terms that depend on the choice of the test matrix $\Omega$ are
$\|\Omega_2\|_2^2$ and $\|\Omega_1^{\dag}\|_2^2$.

We know that the SCM $\Omega$ satisfies the subspace embedding property for the respective 
dual distances and lengths mentioned in the Theorem \ref{theo:BCH}.
This also ensures that the spectral norm of $\Omega_1^{\dag}$ is under
control. We have the condition $\|V_k^\top\Om\Om^\top V_k-I\|_2\leq\eps_0$, implying
\[
 \sqrt{1-\eps_0}\leq\sigma_k(V_k^\top\Om)\leq\sigma_1(V_k^\top\Om)\leq\sqrt{1+\eps_0}.
\]
Then from Lemma 3.6 in \cite{woolfe2008fast}, we have
  \[
  \|\Omega_1^{\dag}\|_2^2=\frac{1}{\sigma_{k}^2(\Omega_1)}\leq
\frac{1}{(1-\eps_0)}.
 \]
 In Lemma \ref{lemm:2cond}, we chose $\eps_0=1/3$ to prove the $(1+\eps)$ approximation. So, we have
 \[
  \|\Omega_1^{\dag}\|_2^2\leq3/2.
 \]

 Next, we bound the spectral norm of $\Omega_2$ as follows
 $ 
\|\Omega_2\|_2^2=\|V_2^\top \Omega\|_2^2\leq\|V_2\|_2^2\|\Omega\|_2^2=\|\Omega\|_2^2=\sigma_1^
2(\Omega),$
 since $V_2$ is an orthonormal matrix. So, we need an upper bound on the top singular value of 
 SCM $\Om$, which we derive from the 
following two lemmas. The first lemma
shows that  
if a code has dual distance $\geq3$, the resulting code matrix $\Phi$  has
orthonormal columns.
 \begin{lemma}[Code matrix with orthonormal columns]\label{lemm:orth}
 A code matrix $\Phi$ generated by a coding scheme which results in codes that
have dual distance $\geq3$, has orthonormal columns. 
\end{lemma}
\begin{IEEEproof}
If a code has dual distance $3$, then the corresponding code matrix (stacked up
codewords as rows) is an 
orthogonal array of strength $2$ \cite{delsarte1998association}. This means all
the tuples of bits,
i.e., $\{0,0\},\{0,1\},\{1,0\},\{1,1\}$, appear with equal frequencies in any
two columns 
of the codeword matrix $C$. As a result, the Hamming distance between any two
columns of $C$ is exactly
$2^{r -1}$ (half the length of the column). This means after the BPSK mapping,
the inner product between any two codewords will be zero. It is easy to see that
the columns are unit norm as well.
\end{IEEEproof}

If there is no downsampling in $\Om$, then the singular values of $\Om$ will simply be $\sqrt{n/\ell}$,
 due to the scaling in \eqref{eq:SCM} of the orthonormal matrix and since $r=\log_2n$. 
 If we downsample the rows of $\Phi$ to form $\Om$, then
the above fact helps us  use  Lemma 3.4 from  \cite{tropp2011improved} which
shows that  
randomly sampling the rows of  a matrix with orthonormal columns results in a 
well-conditioned matrix, and gives bounds for the singular values.
The following lemma is a modification of Lemma 3.4 in \cite{tropp2011improved}. 
\begin{lemma}[Row sampling]\label{lemm:rowsamp}
Let $\Phi$ be a $2^r\times \ell$ code matrix with orthonormal columns and 
let $$M=2^r.\max_{j=1,\ldots,2^r}\|e_j^\top \Phi\|_2^2.$$
  For a positive parameter $\alpha$, select the sample size 
  \[
   n\geq\alpha M\log(\ell).
  \]
  Draw a random subset $T$ from $\{1,\ldots,2^r\}$ by 
  sampling $n$ coordinates without replacement. Then
\begin{equation}
 \sqrt{\frac{(1-\nu)n}{2^r}}\leq\sigma_{\ell}(S_T\Phi) \text{ and } 
 \sigma_{1}(S_T\Phi)\leq \sqrt{\frac{(1+\eta)n}{2^r}}
\end{equation}
with failure probability at most
\[
 \ell.\left[\frac{e^{-\nu}}{(1-\nu)^{(1-\nu)}}\right]^{\alpha
\log(\ell)}+ 
 \ell.\left[\frac{e^{\eta}}{(1+\eta)^{(1+\eta)}}\right]^{\alpha \log(\ell)},
\]
where $\nu\in[0,1)$ and $\eta>0$.
 \end{lemma}
 The  bounds on the singular values of the above lemma are proved in 
\cite{tropp2011improved} 
using the matrix Chernoff bounds.

  Since $n$ is fixed and $M= \ell$ for  code matrices (all the entries of the
matrix are 
$\pm2^{-r/2}$), we get the condition $n\geq \alpha\ell\log(\ell)$.
So, $\alpha$ is less than the ratio $n/\ell\log(\ell)$ and this ratio is typically more than $10$
in the low rank approximation applications.
For $\alpha=10$, we choose $\nu=0.6$ and $\eta=1$, then the failure probability is at most $2\ell^{-1}$.
Since we use the scaling $\sqrt{\frac{2^r}{\ell}}$, the  bounds on the singular
values of 
the subsampled code matrix $\Omega$ will be 
\begin{equation}\label{eq:sigbound}
 \sqrt{\frac{2n}{5\ell}}\leq\sigma_{\ell}(\Omega) \text{ and }
 \sigma_{1}(\Omega)\leq\sqrt{\frac{2n}{\ell}}.
\end{equation}
Thus, we obtain $\|\Omega_2\|_2^2\|\Omega_1^{\dag}\|_2^2=3n/\ell$.
 We substitute this value in \eqref{eq:errorbound} to get the spectral norm error bounds in 
\eqref{eq:error2}.
\end{IEEEproof}

Similarly, we obtain the bounds on the singular values given in \eqref{eq:singerror}  
by substituting the above value of $\|\Omega_2\|_2^2\|\Omega_1^{\dag}\|_2^2$ 
in \eqref{eq:errorbound2} of Theorem \ref{theo:2}.

We observe that the upper bounds for the spectral norm error obtained in
\eqref{eq:error2}  for the SCM is similar to the bounds obtained for 
Gaussian random matrices and structured random matrices like SRFT/SRHT given in 
the review article by Halko et.al \cite{review}.
For the structured random matrices, $(1+\eps)$ optimal Frobenius norm error 
has been derived in \cite{nguyen2009fast} and \cite{boutsidis2013improved}.
We have a similar $(1+\eps)$ optimal Frobenius norm error obtained for subsampled code matrices with
dual distance $>4$ in \eqref{eq:error1}. Importantly, we show that
this optimal error bound can be
achieved with number of samples $\ell=O(k/\eps)$ as opposed to $O(k\log k/\eps)$
required for structured random matrices when the dual distance of the code is $>k$. 
Details on how to generate such 
code matrices with dual distance $>k$ and length $\ell=O(k/\eps)$ is given in section 
\ref{sec:choice}.

\subsection{Differences in the construction}\label{sec:diff}
An important difference between the construction of subsampled code matrices SCM
given in (\ref{eq:SCM}) and
the construction of SRHT or SRFT given in \cite{review,tropp2011improved} is in
the way these matrices are
subsampled.
In the case of SRHT, a Hadamard matrix of size $n\times n$ is applied to input matrix $A$
and $\ell$ out of $n$ columns 
are sampled at random ($n$ must be a power of $2$). When the input matrix is distributively stored, 
this procedure introduces communication issues. The subsampling will require additional
communication 
since each of the nodes must sample the same columns (also recall that we need to sample 
$O(k\log k)$ columns),
making the fast transform slow. Similar issues arise when the input matrices are streamed.

In contrast, in the case of SCM, a $2^r\times \ell$ code
matrix generated from 
an $[\ell,r]$-linear coding scheme 
is considered, and $n$ out of  $2^r$ codewords are chosen (if $r>\log_2n$).
When the input matrix is distributively stored, different rows/columns of the matrix can be sampled by
different codewords locally and hence communicating only the sampled rows/columns.
Similarly, when the input matrix is streamed, at a given time instant, 
the newly arrived rows of the matrix can  simply be sampled by 
new codewords of the code, requiring minimal storage and communication.
For details on space required and communication complexity for sketching streaming
matrices using random sign matrices, see \cite{clarkson2009numerical}.
The subsampling will not affect the $k$-wise independent property of the
code matrix or the distinctness of rows when uniformly subsampled.
This need not be true in the case of SRHT. 
The importance of the distinctness of rows  is discussed next. 

\subsection{Logarithmic factor}\label{sec:logfactor}
A crucial advantage  of the code matrices is that they have very low {\em
coherence}. Coherence is
defined as the maximum inner product between any two rows.
This is in particular true when the minimum distance of the code is close to
half the length. If the 
minimum distance of the code is $d$ then the code matrix generated from an
$[\ell,r]$-code has coherence equal to
$\frac{\ell -2d}{2^r}$. For example, if we consider dual BCH code (see
sec.~\ref{sec:ECC}) the coherence 
is $\frac{2(t-1)\sqrt{\ell+1}-1}{2^r}.$
Low coherence ensures near orthogonality of rows. This is a desirable property
in many applications
such as compressed sensing and sparse recovery.

For a rank-$k$ approximation using subsampled Fourier or Hadamard matrices, we
need to sample $O(k\log k)$ columns. 
This logarithmic factor emerges as a necessary condition 
in the theoretical proof (given in \cite{tropp2011improved}) 
that shows that 
these matrices approximately preserve the geometry of an entire subspace of
input vectors (satisfy the subspace embedding property).
The log factor is also necessary to handle the worst case input matrices. 
The discussions in sec. 11 of \cite{review} and sec. 3.3 of
\cite{tropp2011improved} give more details.
In the case of certain subsampled code matrices,  the log factor is not
 necessary to tackle these worst case input matrices.
To see why this is true, let us consider the worst case example for orthonormal
matrix $V$ described
in Remark 11.2 of \cite{review}.

An infinite family of worst case examples of the matrix $V$ is as follows. For a
fixed integer $k$, let $n=k^2$. 
Form an $n\times k$ orthonormal matrix $V$ by regular decimation of the $n\times
n$ identity matrix. 
That is, $V$ is a matrix whose $j$th row has a unit entry in column $1+(j-1)/k$
when $j\equiv1$ (mod $k$) 
and is zero otherwise. This type of matrix is troublesome when DFT or Hadamard
matrices are used for sampling.

Suppose that we apply $\Omega=DFR^\top $ to the matrix $V^\top $, where  $D$ is same as
in (\ref{eq:SCM}), 
$F$ is an $n\times n$ DFT or Hadamard matrix and $R$ is $\ell\times n$ matrix
that samples $\ell$ 
coordinates from $n$ uniformly at random.
We obtain a matrix $X=V^\top \Omega=WR^\top $, which consists of $\ell$ random columns
sampled from $W=V^\top DF$. 
Up to scaling and modulation of columns, $W$ consists of $k$ copies of a
$k\times k$ DFT or Hadamard matrix
concatenated horizontally.
To ensure that $X$ is well conditioned, we need
$\sigma_k(X)>0$.
That is, we must pick at least one copy of each of the $k$ distinct columns of
$W$. This is the coupon collector's
problem \cite{motwani1995randomized} in disguise
and to obtain a complete set of $k$ columns with non-negligible probability, we
must draw at least
$k\log(k)$ columns.

In the case of code matrices,  we apply a subsampled code matrix $\Omega=DS\Phi$
to the matrix $V^\top $.
We obtain $X=V^\top \Omega=V^\top DS\Phi$, which 
consists of $k$ randomly selected rows of the code matrix $\Phi$. That is, $X$ 
consists of $k$ distinct 
codewords of length $\ell$.
The code matrix has low coherence and all rows are distinct.
If we use a code matrix with dual distance $>k$, then 
 $X$ contains $k$ rows which are $k$-wise independent (near orthonormal)  and
$\sigma_k(X)>0$;  as a result 
the geometry of $V$ is preserved  and
the log factor  is not necessary.
 Thus, for the worst case scenarios we have an $O(\log k)$ factor improvement
over other structured matrices.
 More importantly, this 
 shows that the
 order optimal can be achieved with the immediate lower bound of $O(k)$
 in the number of samples required for the sampling matrices constructed from  deterministic matrices. 
   
 \section{Least squares regression problem}\label{sec:l2reg}
 In this section, we extend the framework to solve the least squares $(\ell_2)$
 regression problem. As discussed in the introduction, 
 the idea of randomized approximations is to 
reduce the dimensions of $A\in\reals^{n\times d}$ and $b\in\reals^{n}$ with $n\gg d$, 
by pre-multiplying them 
by a sampling matrix $\Om\in\reals^{n\times\ell}$, and then to solve the smaller 
problem quickly,
\begin{equation}\label{eq:l2app}
 \min_x\|\Om^\top Ax-\Om^\top b\|_2.
\end{equation}
Let the optimal solution be $x'=(\Om^\top A)^\dag\Om^\top b$.
Here we analyze the performance of SCM as the sampling matrix $\Om$.
We require the sampling matrix $\Om$ to satisfy the JLT and the subspace embedding properties,
which are indeed satisfied by any SCM with dual distance $>4$.
Hence, we can use the results developed by Sarlos \cite{sarlos2006improved}, and Clarkson and
Woodruff \cite{clarkson2009numerical} for our analysis.

 We know that, any code matrix with dual distance $>4$ satisfies the JLT property from our analysis in 
 section \ref{sec:preserve}.
 Sarlos  in \cite{sarlos2006improved}  derived the relation between the 
 JLT matrices and the sampling matrices in the  $\ell_2$
 regression problem  \eqref{eq:l2app}. The following theorem is a modification 
 of theorem 12 in  \cite{sarlos2006improved}.
 \begin{theorem}
  Suppose $A\in\reals^{n\times d},b\in\reals^{n}$.
  Let $\cZ=\min_x\|Ax-b\|_2=\|A\hat{x}- b\|_2$, where $\hat{x}=A^\dag b$ is the minimizer.
  Let $0<\eps,\delta<1$ and $\Om\in\reals^{n\times\ell}$ be a random matrix satisfying JLT and
  $\tilde{\cZ}= \min_x\|\Om^\top (Ax- b)\|_2= \|\Om^\top (Ax'-b)\|_2$, 
  where $x'=(\Om^\top A)^\dag\Om^\top b$. Then, with probability  at least $1-\delta$, we have
  \begin{itemize}
   \item If $\ell=O(\log(1/\delta)/\eps^2)$,
   \begin{equation}
    \tilde{\cZ}\leq(1+\eps)\cZ.
   \end{equation}
     \item If $\ell=O(d\log d.\log(1/\delta)/\eps)$,
   \begin{equation}
    \|Ax'-b\|_2\leq(1+\eps)\cZ.
   \end{equation}
    \item If $\ell=O(d\log d.\log(1/\delta)/\eps^2)$,
   \begin{equation}\label{eq:l3}
    \|\hat{x}-x'\|_2\leq\frac{\eps}{\sigma_{\min}(A)}\cZ.
   \end{equation}
  \end{itemize}
 \end{theorem}
The proof for this theorem can be seen in  \cite{sarlos2006improved}.

If $\sqrt{\|b\|_2^2-\cZ^2}\geq\gamma\|b\|_2$ for some $0<\gamma\leq1$, then we can replace
the last equation \eqref{eq:l3} by 
   \begin{equation}\label{eq:l4}
    \|\hat{x}-x'\|_2\leq\eps\left(\kappa(A)\sqrt{\gamma^{-2}-1}\right)\|\hat{x}\|_2,
   \end{equation}
   where $\kappa(A)$ is the 2~norm condition number of $A$.
(This equation is given to be consistent with the results given in 
the related literature \cite{drineas2006sampling,drineas2011faster,yang2015implementing}.)
Thus, any code matrix with dual distance $>4$ can be used as the sampling matrix 
in the least squares regression problem. Again, 
the performance of such code matrices is very similar to that of
structured random matrices (SRHT) given in \cite{drineas2011faster,boutsidis2013improved}.
Fast multiplication can be used to sample dense input matrix $A$.

Similar to the earlier analysis, we can expect  improved performance when SCM with 
dual distance $>k$ are used
($k$-wise independence property of codes).
For this, we use the bounds derived by Clarkson and Woodruff \cite{clarkson2009numerical}
for random sign matrices. The following theorem which is a modification 
of Theorem 3.1 in  \cite{clarkson2009numerical}
gives the upper bound 
for the regression problem in such cases.
\begin{theorem}
 Given $\eps,\delta>0$, suppose $A\in\reals^{n\times d},b\in\reals^{n}$ and 
 $A$ has rank at most $k$. If $\Om$ is a 
 $\rho(k+\log(1/\delta))$-wise independent with an absolute constant 
 $\rho>1$, and $x'$ and $\hat{x}$ are solutions as defined before, then 
 for $\ell=O(k\log(1/\delta)/\eps)$, with probability at least $1-\delta$, we have
 \[
  \|Ax'-b\|_2\leq(1+\eps)\|A\hat{x}-b\|_2.
 \]
\end{theorem}
This theorem shows that, if the code matrix is 
$\lceil k+\log(1/\delta)\rceil$-wise independent
(i.e., dual distance $>\lceil k+\log(1/\delta)\rceil$),
we can get $\eps-$approximate solution for the regression problem with 
$\ell=O(k\log(1/\delta)/\eps)$ samples. Thus, for the regression problem too, 
we have the $\log k$ factor gain 
in the number of samples over other structured random matrices 
 (SRHT) given in \cite{drineas2011faster,boutsidis2013improved}.

Typically, $A$ is full rank. So,
we will need a code matrix with dual distance $>d$ and 
length of the code $\ell=O(d\log(1/\delta)/\eps)$.
Article \cite{yang2015implementing} discusses the applications where 
such overdetermined system of equations are encountered. In typical applications,
$n$ will be in the range of $10^6-10^9$ and $d$ in the range of $10^2-10^3$
(details in \cite{yang2015implementing}).
In the next section, we discuss how to generate such code matrices with
dual distance $>d$ and minimum length $\ell$, and discuss 
the choice of the error correcting codes
for different type of input matrices and computational environments.

 \section{Choice of error correcting codes}\label{sec:choice}

\subsection{Codes with dual-distance at least $k+1$}
The requirement of $k$-wise independence of codewords translates to the dual distance of the code
being greater than $k$.  Since a smaller code (less number of codewords, 
i.e., smaller $r$) leads to less randomness
in sampling, we would like to use the smallest code with dual distance greater than $k$.

One of the choices of the code can be the family of dual BCH codes. As mentioned earlier,
this family has length $\ell$, dimension $t\log(\ell+1)$ and dual distance at least $2t+1$. 
Hence, to guarantee
dual distance at least $k$, the size of the code must be 
$2^{\frac{k\log(\ell+1)}{2}} = (\ell+1)^{k/2}$.
We can choose $n$  vectors of  length $\frac{k\log(\ell+1)}{2}$ and form the 
codewords by 
 simply multiplying these  with the generator matrix (over $\ff_2$) to form 
 the subsampled code matrix.
 Therefore, forming these code matrices will be much faster than generating 
 $n\times \ell$ i.i.d Gaussian random matrices or 
 random sign matrices which have $k$-wise independent rows.
 
 In general, from the Gilbert-Varshamov bound of coding theory 
 \cite{macwilliams1977theory}, it is known that linear codes
 of size $\sim \sum_{i=0}^{k}\binom{\ell}{i}$ exist that have 
 length $\ell$ and dual distance greater than $k$. The construction of these
 code families are still randomized. However,
when $k = O(\ell)$, or the dual distance is linearly growing with the code length,
the above construction of dual BCH code does not hold in general. 
Infinite families of codes that have distance proportional to the length 
are called {\em asymptotically good codes}.
The Gilbert-Varshamov bound implies that
 asymptotically good linear codes of size $\sim 2^{\ell h(\frac{k}{\ell})}$ 
 exist\footnote{$h(x) \equiv -x\log_2x -(1-x)\log_2 (1-x)$ is the binary entropy function},
 that have length $\ell$ and
dual distance greater than $k$.


\subsection{Choice of the code matrices}
 
Depending on the types of input matrices and the computational environments,
  we can choose different types of code matrices that best suit the applications.
 If the input matrix is a general dense matrix which can be stored in the fast memory (RAM),
 we can choose any structured code matrix with dual distance $>4$, $r=\lceil\log_2 n\rceil$
 (or choose message vectors to be $1$ to $2^{\lceil\log_2 n\rceil}$)
 and $\ell=O(k\log k)$ (eg., dual BCH codes), so that the fast multiplication technique can be 
 exploited (the log factor will not be an issue). This will be similar to using any other
 structured random matrices like SRFT or SRHT. In fact,
 Hadamard matrices are also a class of linear codes, with variants
known as Hadamard codes, Simplex 
  codes or 1st-order Reed-Muller codes. The dual distance of
  Hadamard code is $3$.
  However, with code matrices (say dual BCH codes), subsampling of columns is not 
required, thus reducing randomness and cost.

If the input matrix is sparse and/or is distributively stored, and for parallel 
implementation, we can choose a code matrix with dual distance $>k$ and 
generate them as mentioned earlier and as in section \ref{sec:construct}.
These code matrices are not structured and we can treat them as 
dense transforms (any random matrices), a method to sample such distributively stored matrices
was described  in sec.~\ref{sec:diff}. For SRFT/SRHT sampling matrices, we need
to communicate $O(k\log k)$ columns, but for code matrices with dual distance $>k$, 
the log factor is not necessary.
This will help us overcome the issues with SRFT/SRHT 
 for sparse input matrices and in parallel and distributed applications.
These code matrices are easy to generate (than i.i.d Gaussian random matrices),
the log factor in the number of samples
is not necessary, and thus, using  code matrices in these applications
will reduce randomness and cost  significantly.
When using code matrices, we also have computations gains in the cost of generating 
the sampling matrices, since the code matrices are deterministic, and also  require lower number of 
random numbers to be generated.

A strategy to sample streaming data was also described in sec.~\ref{sec:diff} that requires minimal
storage and communication. For details on the cost, space required and communication complexity 
for sketching streaming
matrices using random sign matrices, we refer \cite{clarkson2009numerical} (observe that 
the SCM matrices are equivalent to random 
sign matrices without the scaling $1/\sqrt{\ell}$).
If the log factor is not an issue (for smaller $k$), then we can choose any code matrix with 
dual distance $>4$ and $r=\lceil\log_2 n\rceil$, and form 
$Y=A\Om$ as a dense transform. These code matrices are 
almost deterministic and unlike SRFT/SRHT,  
subsampling of columns is not required.

In practice, code matrices generated by any linear coding scheme can be
used in place of Gaussian random  matrices.
  As there are many available classes of algebraic and combinatorial codes, we
have a large pool of candidate
  matrices. In this paper we chose dual BCH codes for our numerical experiments
  as they particularly have low
coherence, and turn out to perform quite well in practice.

\begin{table*}[t]
\caption{Classes of sampling matrices with subspace embedding properties}
\label{table:table0}
\begin{center}
\begin{small}
\begin{sc}
\vskip -0.1in
\begin{tabular}{|l|c|c|c|}
\hline
Matrix Classes
&$\ell$ & Runtime &Randomness \\
\hline
Gaussian (or random sign) &$O(k/\eps^2)$&$O(mn\ell)$&$n\ell$\\
SRFT/SRHT~\cite{tropp2011improved,drineas2011faster} &$O(k\log(kn)\log(k/\eps^2)/\eps^2)$
&$O(mn\log\ell)$&$\Theta(n)$\\
Count Sketch~\cite{clarkson2013low}&$(k^2+k)/\eps^2$&$O(\nnz(A))$&$n$\\
Code matrix (dual distance$\geq4$) &$O(k\log(k/\eps)/\eps^2)$&$O(mn\log\ell)$&$n$\\
Code matrix (dual distance$\geq k$)&$O(k/\eps^2)$&$O(mn\log\ell)$&$\Theta(n)$\\
\hline
\end{tabular}
\end{sc}
\end{small}
\end{center}
\vskip -0.1in
\end{table*}
\subsection{Summary of the classes of sampling matrices}
For  readers'  convenience,  we
summarize  in  Table~\ref{table:table0}  a  list  of  some of the classes of  
optimal sampling   matrices which satisfy the subspace embedding property.
The table lists the sampling complexity $\ell$ required for
achieving the $(1+\eps)$ optimal bounds and the runtime cost for 
computing the matrix product $Y=A\Om$. 
The table also lists the amount of random numbers (randomness) required for
each of sampling matrices.
A comprehensive list with 
systematic description of  these and more classes of sampling matrices,
expect  the  last  two  classes can  be  found  in~\cite{yang2015implementing}.


%

\section{Numerical Experiments}\label{sec:results}
The following experiments will illustrate the performance of
 subsampled code matrices as sampling matrices in algorithm \ref{alg:algo1}.
 We compare the performance of dual BCH code matrices against the performance 
 of random Gaussian matrices and subsampled Fourier
transform (SRFT) matrices for different input matrices from various applications.

 Our first experiment is with
 a $4770\times 4770$ matrix named Kohonen from the
 Pajek network (a directed graph's matrix representation), available from
 the UFL  Sparse Matrix Collection \cite{davis2011university}.
Such graph Laplacian matrices are commonly encountered in machine learning and
image processing applications.
The performance of the dual BCH code matrix, Gaussian matrix, subsampled Fourier
transform (SRFT) and Hadamard
(SRHT) matrices are compared as sampling matrices $\Omega$ in algorithm 
\ref{alg:algo1}.   
For SRHT, we have to subsample the rows as well (similar to code matrices) since
the input size is not a power of 2.
All experiments were implemented in matlab v8.1, on an Intel I-5 3.6GHz processor.

\begin{figure*}[!tbh]
\vskip -0.1in
\begin{center}
\begin{tabular}{cc}
\includegraphics[width=0.35\linewidth]{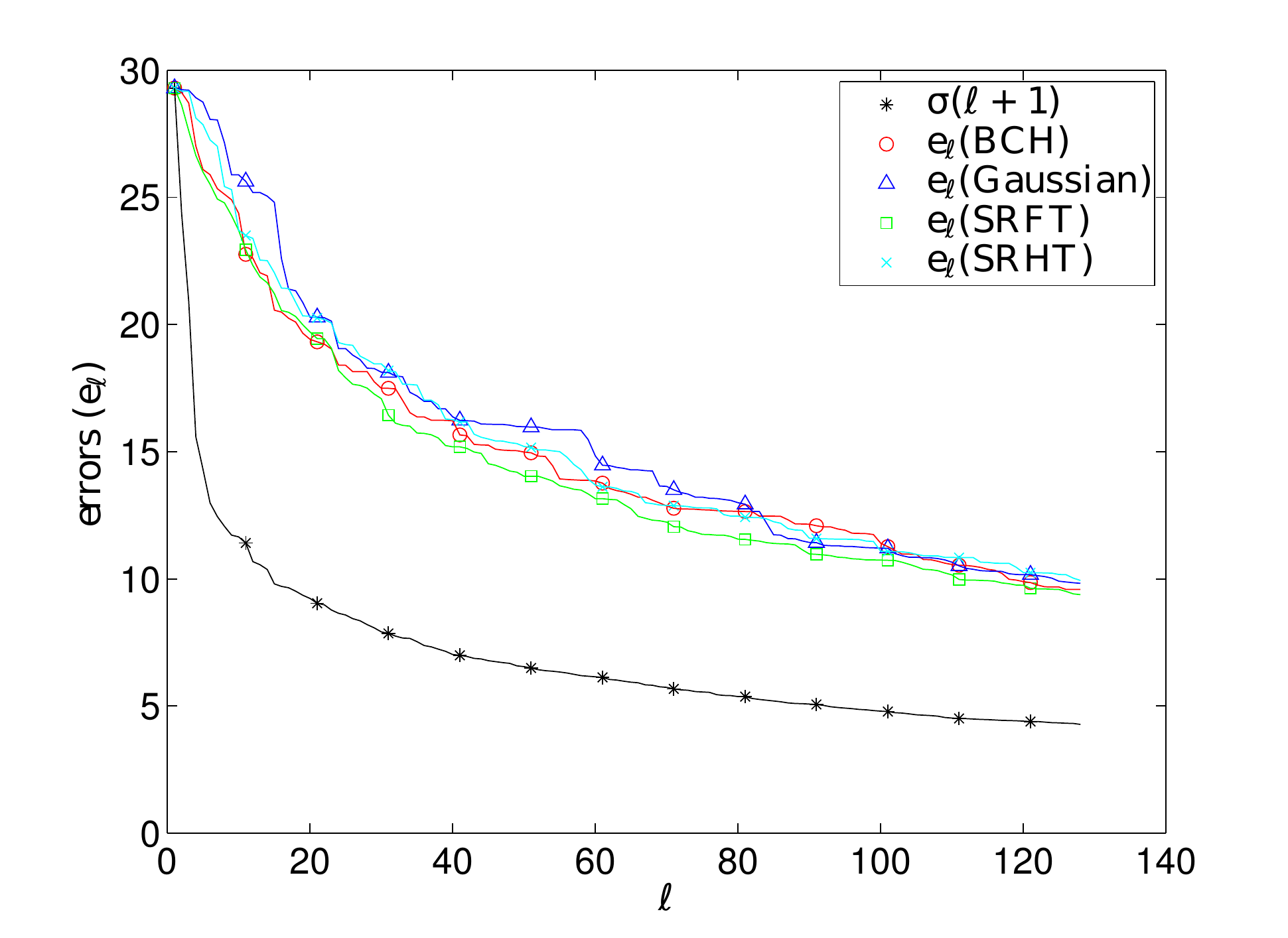} &
\includegraphics[width=0.354\linewidth]{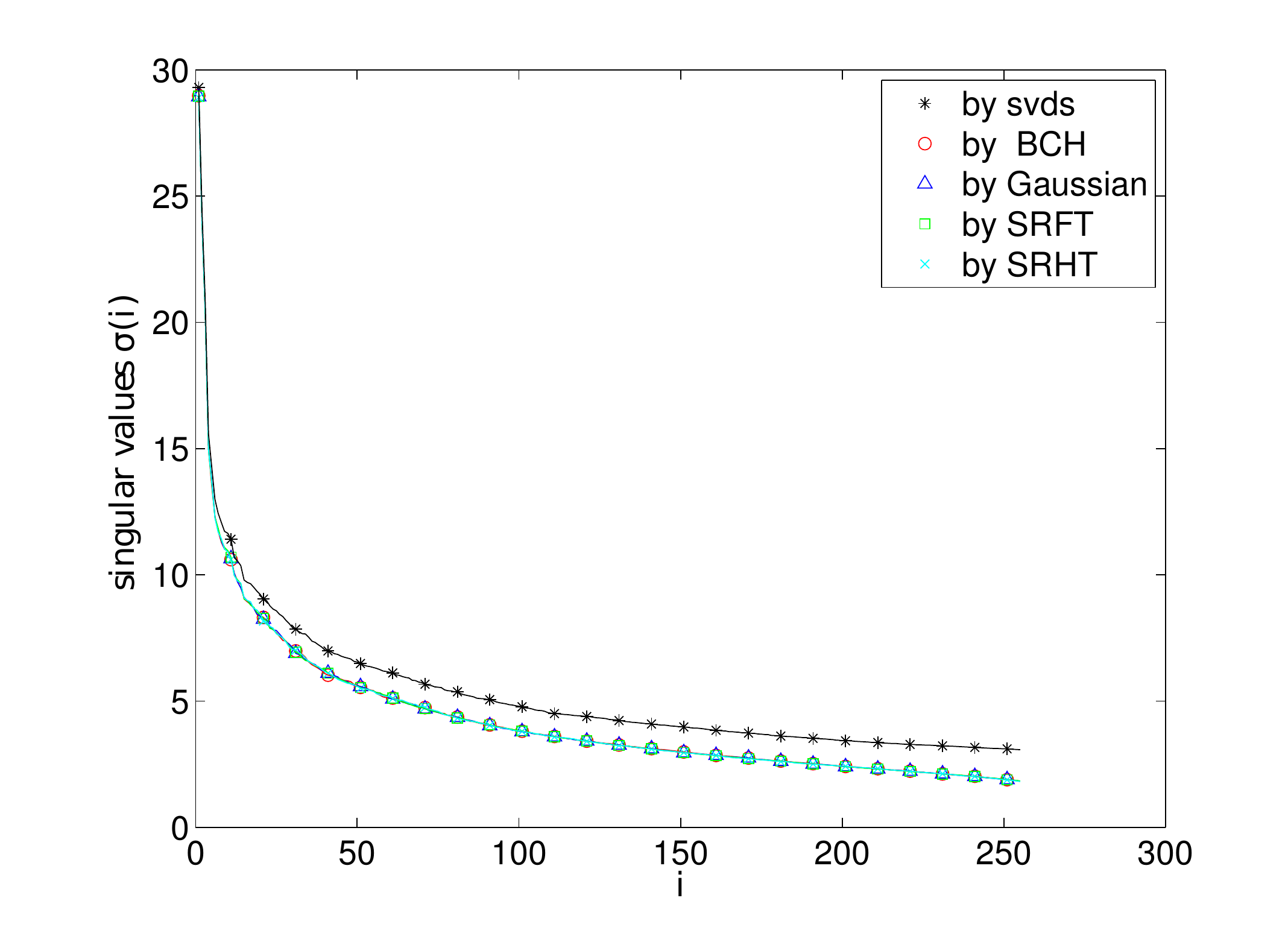} \\
(A)&(B)\\
\end{tabular}
\end{center}
\vskip -0.15in
\caption{ (A) The theoretical minimum $\sigma_{\ell+1}$ and approximate error as a
 function of the number of 
 random samples $\ell$  using  dual BCH code, Gaussian, SRFT and SRHT
matrices as sampling matrix in Algorithm \ref{alg:algo1} for input matrix 
 \texttt{Kohonen}.
(B) Estimates for top 255  singular values computed by Algorithm \ref{alg:algo1} using
dual BCH code, Gaussian,
SRFT and SRHT matrices and the exact singular values by svds
function.\label{fig:fig1}}
\vskip -0.1in
\end{figure*}

\begin{table*}[t]
\caption{Comparison of errors}
\label{table:table1}
\begin{center}
\begin{small}
\begin{sc}
\vskip -0.1in
\begin{tabular}{|l|c|c|c|c|c|}
\hline
& && Dual BCH&
Gaussian&
SRFT\\\cline{4-6}
Matrices&Sizes&$\ell$&Error &Error &Error\\

\hline
lpiceria3d&$4400\times3576$ &63&$16.61$&$17.55$ &$17.32$\\
Delaunay &$4096\times4096$&63&$6.386$&$6.398$& $6.383$\\
deter3&$21777\times7647$ &127&$9.260$&$9.266$& $9.298$\\
 EPA&$4772\times4772$&255&$5.552$&$5.587$&$5.409$\\
Kohonen&$4770\times 4770$&511&$4.297$&$4.294$&$4.261$\\
\hline
ukerbe1&$5981\times5981$&127&$3.093$&$ 3.0945$& $3.092$\\
dw4096&$8192\times8192$&127&$108.96$&$ 108.93$& $108.98$\\
FA&$10617\times10617$&127&$2.19$&$2.17$& $2.16$\\
qpband&$20000\times20000$&63&$4.29$&$4.30$& $4.26$\\
\hline
\end{tabular}
\end{sc}
\end{small}
\end{center}
\vskip -0.1in
\end{table*}

Figure \ref{fig:fig1}(A) gives the actual error
$e_{\ell}=\|A-Q^{(\ell)}(Q^{(\ell)})^\top A\|$ for each $\ell$ number 
of samples 
when a subsampled dual BCH code matrix, a Gaussian matrix, SRFT and SRHT
matrices are used as sampling matrices in algorithm \ref{alg:algo1},
respectively.  
The best rank-$\ell$ approximation error $\sigma_{\ell+1}$ is also given. 
Figure \ref{fig:fig1}(B) plots the singular values obtained from algorithm
\ref{alg:algo1}, for $\ell=255$ and
different sampling matrices $\Omega$ used.
The top $255$ exact singular values of the matrix (available in the UFL
database) are also plotted.
We observe that, in practice, the performance of all four sampling matrices are similar.

Table \ref{table:table1} compares the errors $e_{\ell}$ for $\ell$ number of
samples, obtained for 
a variety of input matrices from different applications when subsampled dual BCH
code, Gaussian and
SRFT matrices were used.
All  matrices were obtained from the UFL database~\cite{davis2011university}. 
Matrices lpi\_ceria3d
 and
deter3 are from linear programming problems. 
S80PI\_n1  and dw4096 are from an eigenvalue/model reduction problem.
 Delaunay, EPA,  ukerbe1, FA (network) and Kohonen are 
graph Laplacian matrices. 
qpband is from an optimization problem.
The table depicts two sets of experiments (divided by the line). 
The first set (top five examples)
illustrates 
how errors vary as the sample size $\ell$ is increased.
The second set (bottom four) illustrates how the errors vary as the
size of the matrix increases. For the last matrix (qpband) we could 
compute the decomposition
for only $\ell=63$ due to memory restrictions.
We observe that, for small $\ell$,  in the first five examples the
error performance of  code matrices is slightly better than that of  Gaussian
matrices. 
For higher $\ell$, the error remains similar to the error for Gaussian matrices.
All input matrices are sparse, hence we cannot use the fast transforms. 
We still see that code matrices take less time than both Gaussian and SRFT.
In practice, we can use code matrices in place of fully random (Gaussian)
 matrices or structured random matrices due to the advantages of code matrices over the other
 sampling matrices, as discussed in the previous sections.
 Next, we illustrate the performance of algorithm \ref{alg:algo1}
 with different sampling matrices in a practical application.

\paragraph*{\bf Eigenfaces} Eigenfaces is a popular method for face recognition
that is based on Principal Component Analysis (PCA)
\cite{turk1991eigenfaces,sirovich2009symmetry}. In this experiment
(chosen as a verifiable comparison with results in \cite{mingu2014}),
we demonstrate the performance of randomized algorithm with different sampling
matrices on face recognition.
The face dataset is obtained from the AT\&T Labs Cambridge database of faces
\cite{att2002}. 
There are ten different images of each of $40$ distinct subjects.
The size of each image is $92\times112$ pixels, with $256$ gray levels per
pixel. 
$200$ of these faces, $5$ from each individual are used as 
training images and the remaining $200$ as test images to classify.

 \begin{table}[t]
\caption{Comparison of the Number of Incorrect Matches}
\label{table:table2}
\begin{center}
\begin{small}
\begin{sc}
\vskip -0.1in
\begin{tabular}{|l|c|c|c|c|}
\hline
Rank & \multicolumn{1}{p{2.0cm}}{\centering Dual BCH \\ $p$} &
\multicolumn{1}{p{1.5cm}}{\centering Gaussian \\ $p$} &
\multicolumn{1}{p{1.2cm}}{\centering SRFT\\ $p$} & T-SVD\\\cline{2-5}
$k$& $10\:\:\:20$& $10\:\:\:20$&$10\:\:\:20$&\\
\hline
$10$ &$18\:\:\:13$&$19\:\:\:15$& $21\:\:\:18$&$26$\\
$20$ &$14\:\:\:11$&$14\:\:\:12$&$16\:\:\:12$&$13$\\
$30$ &$10\:\:\:08$&$13\:\:\:08$&$12\:\:\:09$&$10$\\
$40$  & $09\:\:\:08$ &$08\:\:\:07$&$08\:\:\:10$&$06$\\
\hline
\end{tabular}
\end{sc}
\end{small}
\end{center}
\vskip -0.2in
\end{table}

In the first step, we compute the principal components (dimensionality
reduction)  of mean  shifted  training image  dataset using  algorithm
\ref{alg:algo1}, with different sampling matrix $\Omega$ and different
$p=\ell-k$ values (oversampling used). Next, we project the mean-shifted images into the singular
vector space using the singular  vectors obtained from the first step.
The projections are  called feature vectors and are  used to train the
classifier.   To classify  a new  face,  we mean-shift  the image  and
project it onto the singular  vector space obtained in the first step,
obtaining a new feature vector.
The new feature vector is classified using a classifier  which is trained on the
feature vectors from the training images.
We used the in-built MATLAB function {\btt classify} for feature training and
classification.
We compare the performance of the 
dual BCH code matrix, Gaussian matrix and SRFT matrix against exact 
truncated SVD (T-SVD).
The  results are  summarized in  Table \ref{table:table2}.  For $p=10$
dual BCH code matrices  give results that are similar to those of
 truncated  SVD, and  for rank $k<40$, $p=20$ our  results are 
 superior.

\section{Conclusion}
This paper advocated the use of matrices generated by error correcting
codes   as   an  alternative   to   random   Gaussian  or   subsampled
Fourier/Hadamard    matrices    for    computing   low rank    matrix
approximations.   Among  the  attractive  properties of  the  proposed
approach  are  the  numerous  choices  of  parameters  available,  
ease of generation, reduced randomness and cost, and the near-orthogonality of rows.  
We showed that any code matrices with dual distance $>4$ satisfy the subspace embedding
property, and we can achieve $(1+\eps)$ optimal Frobenius norm error bound. 
Indeed if the dual distance of the code matrix 
is $>k$, then the length of the code (sampling complexity) required is in $O(k)$, thus
leading to an order optimal in 
the worst-case guaranteed sampling complexity, an
 improvement by a factor of $O(\log k)$ 
  over other known almost
deterministic matrices.
We saw that fast multiplication is possible with structured code matrices,
resulting in fast approximations for general dense input matrices.
The implementation issues of FFT-like structured random matrices 
in the  parallel and distributed environments can be overcome by using 
code matrices as sampling matrices.

It is known that Gaussian matrices perform much better in practice compared to
their theoretical
analysis \cite{review}. Our code matrices (a) are almost deterministic, and (b)
have $\pm 1$ entries.
Still, they perform equally well (as illustrated by  experiments) compared to
random real Gaussian matrices and complex Fourier matrices. 
Because of  the availability of different families  of classical codes
in the rich literature of coding theory, many possible choices of code
matrices are at hand.  One of  the contributions of this paper is to
open  up these  options for    use as  structured sampling  operators in
low-rank approximations and least squares regression problem.  

Interesting future works include extending the framework of code matrices
to other similar applications.
The connections between code matrices and JLT and random sign matrices might
lead to improved analysis in other applications of codes such as sparse recovery
\cite{barg2015restricted}.


\end{document}